\RequirePackage{lineno}
\documentclass[aps,floatfix,prc,twocolumn,showpacs,showkeys,amsmath,amssymb,superscriptaddress]{revtex4}

\usepackage{soul}
\usepackage{graphicx}	\usepackage{xspace}     \usepackage{amsmath}
\usepackage{hyperref}\usepackage{color}\usepackage{multirow}\usepackage{booktabs,array}

\makeatletter{}\newcommand{\jhcs}{jet-hadron correlations\xspace}
\newcommand{\dhcs}{di-hadron correlations\xspace}
\newcommand{\Jhcs}{Jet-hadron correlations\xspace}
\newcommand{\Dhcs}{Di-hadron correlations\xspace}

\newcommand{\dhc}{di-hadron correlation\xspace}
\newcommand{\pT}{$p_{\rm T}$\xspace}

\newcommand{\zerodeg}{0$^{\circ}$\xspace}
\newcommand{\oneeightydeg}{180$^{\circ}$\xspace}
\newcommand{\dphi}{$\Delta\phi$\xspace}
\newcommand{\dphirange}[1]{$|\Delta\phi|< $#1\xspace}

\newcommand{\etarange}[1]{$|\eta|<$#1\xspace}
\newcommand{\deta}{$\Delta\eta$\xspace}
\newcommand{\GeV}{GeV/$c$\xspace}

\newcommand{\dNeventdphideta}{$\frac{1}{N_{\mathrm{e}}} \frac{d^2N}{d\Delta\phi d\Delta\eta}$\xspace}
\newcommand{\psiR}{$\psi_{R}$\xspace}
\newcommand{\Beff}{$\tilde{\beta^{R}}$\xspace}
\newcommand{\vn}{$v_{n}$\xspace}
\newcommand{\vneff}{$\tilde{v}_{n}^{t}$\xspace}
\newcommand{\vnum}[1]{$v_{#1}$\xspace}
\newcommand{\vnumtrigger}[1]{$v_{#1}^{t}$\xspace}
\newcommand{\vnumeff}[1]{$\tilde{v}_{#1}$\xspace}
\newcommand{\vnumeffsq}[1]{$\tilde{v}_{#1}^{2}$\xspace}
\newcommand{\vnumeffR}[1]{$\tilde{v}_{#1}^{R,t}$\xspace}
\newcommand{\vnumeffassoc}[1]{$\tilde{v}_{#1}^{a}$\xspace}
\newcommand{\vnumefftrigger}[1]{$\tilde{v}_{#1}^{t}$\xspace}
\newcommand{\vneffR}{$\tilde{v}_{n}^{R,t}$\xspace}
\newcommand{\vnsq}{$\tilde{v}_{n}^{\mathrm{a}}\tilde{v}_{n}^{\mathrm{t}}$\xspace}
\newcommand{\vntrigger}{$v_{n}^{\mathrm{t}}$\xspace}
\newcommand{\vnassoc}{$v_{n}^{\mathrm{a}}$\xspace}
\newcommand{\vnefftrigger}{$\tilde{v}_{n}^{\mathrm{t}}$\xspace}
\newcommand{\vneffassoc}{$\tilde{v}_{n}^{\mathrm{a}}$\xspace}

\newcommand{\Fref}[1]{Figure~\ref{#1}}
\newcommand{\Tref}[1]{Table~\ref{#1}}
\newcommand{\Eref}[1]{Equation~\ref{#1}}
\newcommand{\dNdphi}{$\frac{dN}{d\phi}$}
\newcommand{\pp}{$p$+$p$\xspace}

\newcommand{\Pb}{$Pb+Pb$\xspace}
\newcommand{\dAu}{$d+Au$\xspace}

\newcommand{\AplusA}{$A$+$A$\xspace}
\newcommand{\RAA}{$R_{AA}$\xspace}
\newcommand{\sqrts}{$\sqrt{s}$}
\newcommand{\sNN}{$\sqrt{s_{\mathrm{NN}}}$}
\newcommand{\ns}{near-side\xspace}
\newcommand{\as}{away-side\xspace}

\newcommand{\chisq}[2]{$\chi^2$/NDF = #1/#2}
\newcommand{\ptassoc}{$p_T^{\mathrm{a}}$\xspace}
\newcommand{\pttrig}{$p_T^{\mathrm{t}}$\xspace}

\newcommand{\pttrigrange}[2]{#1 $< p_T^{\mathrm{t}} <$ #2~GeV/$c$\xspace}
\newcommand{\ptassocrange}[2]{#1 $< p_T^{\mathrm{a}} < $ #2~GeV/$c$\xspace}

\newcommand{\Rn}{$r_n$\xspace}
\newcommand{\R}[1]{$r_{#1}$\xspace}
\newcommand{\midplane}{mid-plane\xspace}
\newcommand{\inplane}{in-plane\xspace}
\newcommand{\outplane}{out-of-plane\xspace}
\newcommand{\allplane}{all reaction plane angles\xspace}
\newcommand{\Midplane}{Mid-plane\xspace}
\newcommand{\Inplane}{In-plane\xspace}
\newcommand{\Outplane}{Out-of-plane\xspace}

\begin{document}

\title{Background subtraction methods for precision measurements of di-hadron and jet-hadron correlations in heavy ion collisions}

\author{Natasha Sharma$^{1}$, Joel Mazer$^{1}$, Meghan Stuart$^{1}$, Christine Nattrass} \affiliation{
University of Tennessee, Knoxville, Tennessee 37996, USA. \\
}
\date{\today}

\begin{abstract} 
\Dhcs and \jhcs are frequently used to study interactions of the quark gluon plasma with the hard partons that form jets.  The existing background subtraction methods for these studies depend on several assumptions and independent measurements of the Fourier coefficients of the combinatorial background.  
In this paper, we present a method for determining the background using a fit to the reaction plane dependence of the background-dominated region of the \ns to extract the background.  We also fit the of the background-dominated region of the \ns without the reaction plane dependence.
To test the accuracy of these methods, a simple model is used to simulate di-hadron and \jhcs with a combinatorial background similar to that observed in the data.  The true signal is compared to the extracted signal. The results are compared to results from two variants of the zero-yield-at-minimum (ZYAM) method.  We test these methods for mid-peripheral and central collisions and for di-hadron and \jhcs.  These methods are more precise than the ZYAM method with fewer assumptions about the shape and level of the combinatorial background, even in central collisions where the experimental resolution on the measurement of the reaction plane dependence is poor.  These methods will allow more accurate studies of modifications of the \as jet and will be particularly useful for studies of \jhcs, where the combinatorial background is poorly constrained from previous studies.

\end{abstract}

  \maketitle

 \section{Introduction}\label{Sec:Introduction}
\makeatletter{}The quark gluon plasma (QGP), a strongly interacting liquid of quarks and gluons, is produced in high-energy nuclear collisions~\cite{Adcox:2004mh,Adams:2005dq,Back:2004je,Arsene:2004fa}.  
Hard probes such as jets are frequently used to study the QGP because they are produced by hard scatterings early in the collision and propagate through the medium.
Hard partons interact with the medium and lose energy, a process called jet quenching.

The interactions of jets with the medium are commonly studied using three experimental methods: measurements of single-particle spectra at high \pT; di-hadron correlations where at least one particle is at high momentum, and fully reconstructed jets.  Observations of jet quenching at RHIC was one of the key signatures of the formation of the QGP.  The initial observation relied on measurements of the nuclear modification factor \RAA, which compares the hadron spectrum in \AplusA to that in \pp.  At high momentum where hadron production is expected to be dominated by jets (\pT $>$ 5 \GeV at RHIC), the number of hadrons observed in \AplusA collisions is roughly 1/5 that expected from \pp collisions~\cite{Adcox:2001jp,Adler:2002xw,Adler:2003qi,Back:2004bq,Adams:2003kv}.

Studies of jets in a QGP are complicated by the large background due to soft processes.  Not only is there a large background but the strong collective flow observed in bulk particle production leads to correlations between particles in the bulk similar to the correlations due to jet production.  Since both jets and collective flow contribute to correlations between particles, collective flow generates a significant background for any study of jets in heavy ion collisions.  Collective flow is dominant at low momenta (\pT $\lesssim$ 2 \GeV), so this background has typically been dealt with by focusing studies of jets on high momentum particles.  However, gluon bremsstrahlung leads to gluons that are softer than the parent parton.  As these hadronize, the final-state hadrons are softer on average than the final-state hadrons from the parent parton~\cite{Vitev:2008rz}. This means that many of these modifications are likely to be concentrated at low momentum and at large angles from the parent parton.  Therefore, a reliable and precise method for background subtraction is needed in order to quantify jet modification at low and intermediate momenta.

We extract the background using a fit to the reaction plane dependence of background-dominated region on the \ns of di-hadron correlations and jet-hadron correlations, called the reaction plane fit (RPF) method.  We also compare to a fit without the reaction plane fit, the near-side fit (NSF) method.
We demonstrate that these methods produce more accurate and reliable results than the ZYAM method.  These methods both take advantage of differences between the signal and the background in order to determine the background.  The signal on the \ns is concentrated in a peak near the trigger hadron or jet.  This peak is narrow in both azimuth and pseudorapidity.  In contrast, the background forms a peak in azimuth but not pseudorapidity.  For analyses in a narrow pseudorapidity range, the background is roughly independent of pseudorapidity.  
We combine a background roughly matching that observed in \AplusA collisions with a known signal generated from PYTHIA~\cite{Sjostrand:2006za} to show that these methods accurately and reliably reconstruct the signal.

All background subtraction methods for \dhcs and \jhcs to date have assumed that the shape of the background is known.  Our methods also make this assumption.  Previous methods further assume that the magnitude of the coefficients of the Fourier decomposition of the background are known from other studies and can be measured independent of the correlation.  Our methods do not make this assumption; effects that could modify these coefficients such as contributions from jets or differences in the hydrodynamical flow in events that contain jets, are taken into account because the Fourier coefficients are fit.  Moreover, this allows the accurate determination of the background even in cases where the Fourier coefficients have not been measured to higher order, such as \jhcs.  Like previous methods, our methods are also dependent on the assumption that contributions from other correlations such as Hanbury-Brown-Twiss correlations or decays of resonances that are not part of a jet are negligible.  The ZYAM method makes the assumption that there is an angle in azimuth for which there are no correlations from jets.  At sufficiently low momenta, this is certainly not true, since the \ns and \as peaks overlap in azimuth.  
Our methods do not assume that the signal is zero at a given angle, however, we assume instead that the signal is negligible on the \ns when the difference between the pseudorapidities of the associated particle and the trigger is large.

We first summarize correlation studies, discussing previous studies, correlations that contribute to the background, the shape of the background, and the shape of the signal.  This is a motivation for our model of the background and the signal, discussed in the following section.  We then show the results of the NSF method.  While the NSF method is more accurate than the ZYAM method, the results are not stable when the fit range is reduced.  The RPF method is tested for \dhcs in mid-peripheral collisions with a fit over a wide and a narrow range in azimuth, demonstrating that this method is more robust than the NSF method and produces more precise results than the ZYAM method.  We then test this method for central collisions.  Even though the reaction plane resolution is poor in central collisions, the limited information available constrains the background and produces more precise results than the ZYAM method.  Finally, the method is tested for \jhcs.

\section{Correlation studies}
In a typical \dhc study~\cite{Adler:2002tq,Adler:2005ad,Abelev:2009af,Aamodt:2011vg,Alver:2009id}, a high-\pT trigger particle is identified and used to define the origin in azimuth and pseudorapidity.  Typically, it is defined by its high momentum alone, restricted to a range of momenta.  Here all trigger particles in a given momentum region are accepted and then
the correlation between particles is studied in both azimuth and pseudorapidity.  By selecting high-\pT particles the fraction of trigger particles coming from the production of jets is enhanced, however,  it is not possible to determine conclusively which trigger particles originate from hard processes and which arise from soft processes, particularly for lower-momentum ($<$10 \GeV) trigger particles.

Associated particles are also usually defined only as particles within a given momentum interval.  For each associated particle in the event, its position relative to the trigger particle in azimuth (\dphi) and pseudorapidity (\deta) is determined and the conditional yield is calculated.  In this paper, the conditional yield is normalized by the number of events.  
As with the trigger particle, it is not possible to determine conclusively which associated particles are from hard processes and which are from soft processes.
For \jhcs, instead of a trigger hadron, a jet candidate is used to define the origin in azimuth and pseudorapidity~\cite{Adamczyk:2013jei,CMS-PAS-HIN-14-016}.  Below, ``trigger'' refers to either a trigger particle or a trigger jet.

Since both the associated particles and the trigger include particles created from and modified by soft processes such as hydrodynamical flow, the combinatorial background is not azimuthally isotropic in heavy ion collisions.  The way that this combinatorial background is typically treated is to assume that the contribution to both the trigger and associated particles can be factorized into a contribution from hard processes, the signal ($J$), and a contribution from soft processes, the background ($B$).  This is referred to as the two-source model~\cite{Adler:2005ee}.  The term ``raw signal'' is used below for what would be measured experimentally after corrections for detector and acceptance effects but before background subtraction.  The raw signal contains signal-signal ($J$-$J$) correlations, signal-background ($J$-$B$) correlations, background-signal ($B$-$J$) correlations, and background-background ($B$-$B$) correlations.  It is assumed that the processes that produce the signal and the background are completely independent so that the  $J$-$B$ and $B$-$J$ correlations are also background.

\begin{figure}
\begin{center}
\rotatebox{0}{\resizebox{8cm}{!}{
        \includegraphics{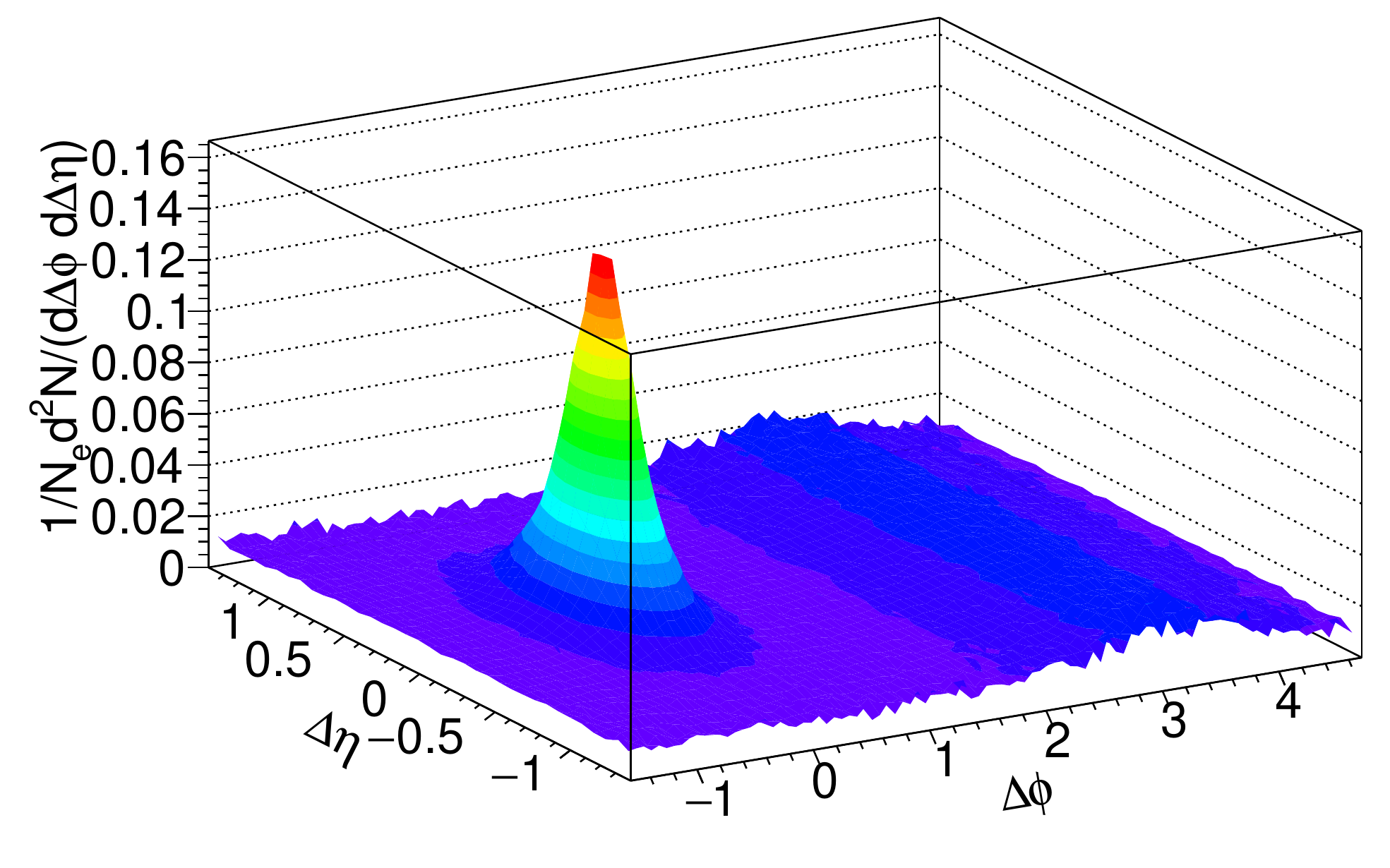}
}}\caption{\Dhcs for trigger momenta \pttrigrange{8}{10} within pseudorapidities \etarange\ {0.5} and associated particles within \etarange\ {0.9} with momenta \ptassocrange{1.0}{2.0} in \pp collisions at \sqrts\ = 2.76 TeV in PYTHIA~\cite{Sjostrand:2006za}.  The signal is normalized by the number of equivalent \Pb collisions in our simulations and corrected for the acceptance using the mixed event correction described in the text.  }\label{Fig:Signal}
\end{center}
\end{figure}

A typical raw signal from \dhcs for trigger momenta \pttrigrange{8}{10} within pseudorapidities \etarange\ {0.5} and associated particles within \etarange\ {0.9} with momenta \ptassocrange{1.0}{2.0} in \pp collisions at \sqrts\ = 2.76 TeV in PYTHIA~\cite{Sjostrand:2006za} is shown in~\Fref{Fig:Signal}.  The raw rsignal is normalized by the number of equivalent \Pb collisions because this is used as the known signal later and added to the background in \Pb collisions.  Because even PYTHIA has background from an underlying event, this raw signal includes $J$-$J$, $J$-$B$, $B$-$J$, and $B$-$B$ correlations.  Because there is no physical correlation between the signal and the background in PYTHIA, the $J$-$B$, $B$-$J$, and $B$-$B$ correlations are independent of azimuth and lead to the plateau in~\Fref{Fig:Signal}.  \Fref{Fig:Signal} shows that there is a peak near \zerodeg which is narrow in both \dphi and \deta.  There is also a peak near \oneeightydeg, which is narrow only in \dphi, however, this peak is roughly independent of pseudorapidity.  The former is called the \ns and comes from associated particles from the same parton as the one that generated the trigger particle.  The latter is called the \as and comes from associated particles from the parton that scattered off of the parton that generated the trigger particle.  

The parton that produces the \ns is generally thought to be biased towards partons that have not interacted strongly with the medium and it is therefore critical to study the \as peak.  Even in PYTHIA the \as peak is roughly independent of pseudorapidity within the typical acceptance used in correlation analyses, as shown in~\Fref{Fig:Signal}.  While a hard parton scattering produces two back-to-back partons in the rest frame of the parton, the rest frame of the parton is, in general, not the same as the rest frame of the incoming nuclei.  The difference in azimuth is negligible since most of the momenta of both the parton and the nuclei are in the direction of the beam pipe, however, the difference in pseudorapidity can be quite substantial.  This causes the \as to be broad in \deta without modified fragmentation or interaction with the medium.  This is evident in~\Fref{Fig:Signal}.

In a heavy ion collision, these peaks may be widened through partonic interactions with the medium~\cite{Armesto:2009ug}, for instance if the original partons have emitted bremsstrahlung gluons.  The yield in the peaks may either be higher or lower, depending on how the parton interacted with the medium and on the specific momentum range.  A parton that emitted a bremsstrahlung gluon would have less energy when it fragments, meaning that the peak at high associated particle momentum would be depleted because the odds of producing a high momentum particle through fragmentation would be lower.  However, at lower momentum, the peak would include particles from fragmentation of both the parent parton and the bremsstrahlung gluon, so the peak would be enhanced at low associated momenta.

For studies of the \ns, the raw signal at large \deta can be used to determine the level of the background without extracting the precise \vn~\cite{Abelev:2009af,Agakishiev:2011st,Abelev:2009ah}.  However, this is only useful for studies of the \ns because the signal on the \as is also roughly independent of \deta and will also be subtracted when this method is applied. 

The background due to soft processes can be written in general by a Fourier decomposition of the azimuthal anisotropy relative to the reaction plane:
\begin{equation}
 \frac{dN}{d(\phi-\psi_{R})} \propto 1 + \sum_{n=1}^{\infty} 2 v_{n} \cos[n(\phi - \psi_{R})],\label{Eqtn:FlowFourierDecomposition}
\end{equation}
\noindent where $N$ is the number of particles, $\phi$ is the angle of a particle's momentum in azimuth in detector coordinates and \psiR is the angle of the reaction plane in detector coordinates.  In high energy heavy ion collisions the Fourier coefficients \vn arise due to hydrodynamical flow~\cite{Adcox:2004mh,Adams:2005dq,Back:2004je,Arsene:2004fa}.  The initial overlap region is azimuthally anisotropic, leading to anisotropic pressure gradients, which give rise to the \vn.  These initial azimuthal anisotropies are preserved through partonic hydrodynamical flow and lead to azimuthal anisotropies in the final-state hadrons.  The magnitude of the Fourier coefficients \vn decreases with increasing order.  The sign of the first-order coefficient $v_{1}$ is dependent on the incoming direction of the nuclei and changes sign when going from positive to negative pseudorapidities.  Since correlation analyses typically average over both positive and negative pseudorapidities, the average $v_1$ is zero.  

The even \vn are generally understood to arise mainly from anisotropies in the average overlap region of the incoming nuclei, considering the nucleons to be smoothly distributed in the nucleus with the density depending only on the radius.  
The \vn with even $n$ are correlated with the reaction plane.  The odd \vn are generally understood to arise from the fluctuations in the positions of the nucleons within the nucleus.  High-energy heavy ion collisions happen on a time scale short enough to be sensitive to the position of individual nucleons within the nucleus.  Since these fluctuations are not causally related to the reaction plane, the odd \vn are assumed to be uncorrelated with the reaction plane. Recent measurements by ATLAS confirm that the correlation between $n$ = 2 and $n$ = 3 reaction planes is very weak~\cite{Aad:2014fla}.

\begin{figure}
\begin{center}
\rotatebox{0}{\resizebox{8cm}{!}{
        \includegraphics{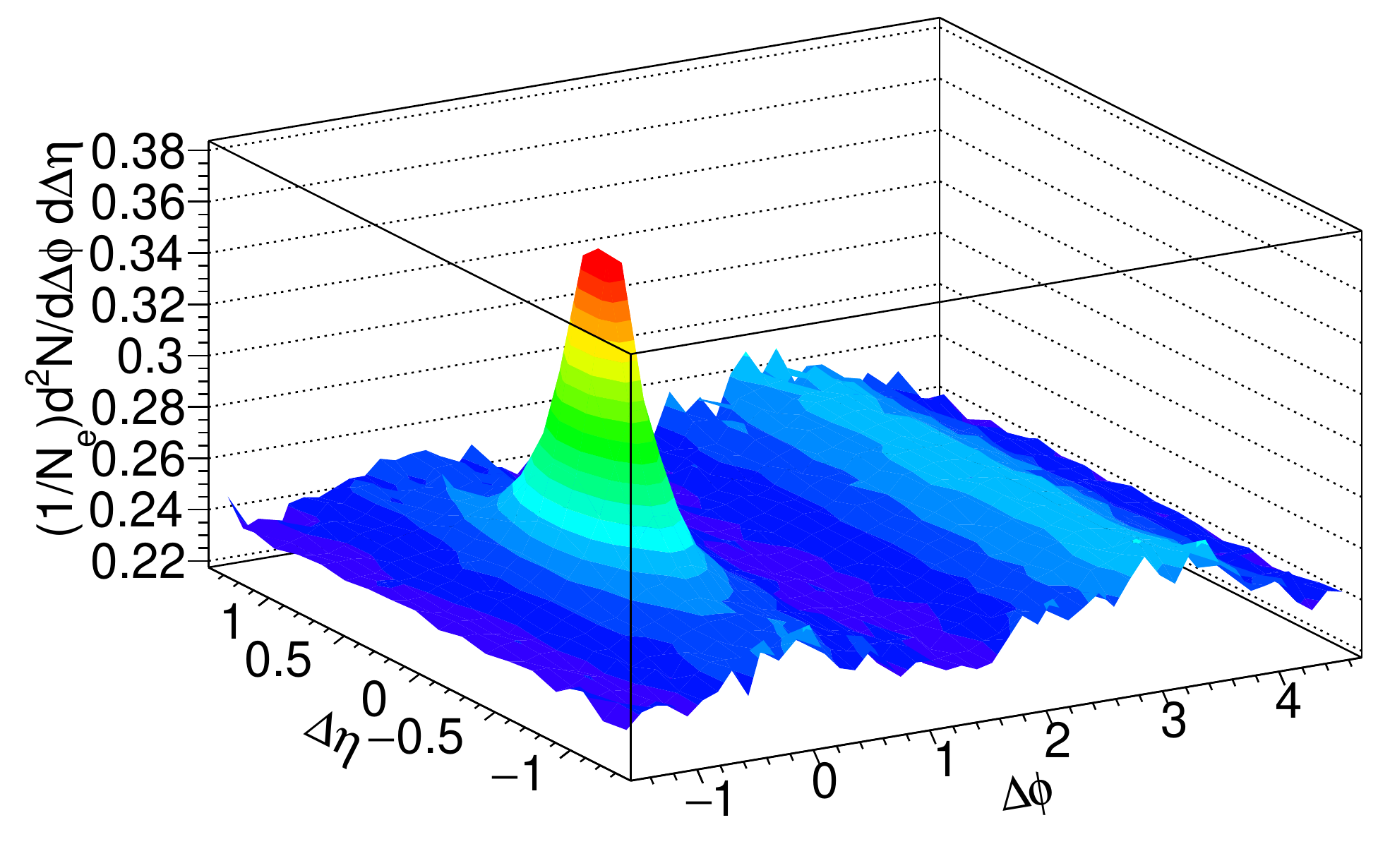}
}}\caption{Di-hadron correlation signal for trigger momenta \mbox{\pttrigrange{8}{10}} within pseudorapidities \etarange\ {0.5} and associated particles within \etarange\ {0.9} with momenta \ptassocrange{1.0}{2.0} in 30-40\% \Pb collisions at \sNN\ = 2.76 TeV.  The signal is normalized by the number of \Pb collisions.  The signal is from \Fref{Fig:Signal} and the generation of the background is discussed in the method section.  }\label{Fig:SignalPlusBackground}
\end{center}
\end{figure}

For $B$-$B$ correlations entirely due to hydrodynamical flow the conditional yield will be given by~\cite{Bielcikova:2003ku}:
\begin{equation}
 \frac{dN}{\pi d\Delta\phi} = B[1 + \sum_{n=1}^{\infty} 2 v_{n}^{\mathrm{t}} v_{n}^{\mathrm{a}} \cos(n\Delta\phi)],\label{Eqtn:BBCorrelations}
\end{equation}
\noindent where $B$ is a constant that depends on the multiplicity of trigger and associated particles in an event and on the normalization convention, \dphi is the difference in azimuthal angle between the associated particle and the trigger, \vntrigger is the \vn for the trigger, and \vnassoc is the \vn for the associated particle.  $B$-$B$ correlations due to processes other than hydrodynamical flow are generally assumed to be negligible.  In a typical analysis, the pseudorapidity range for both trigger and associated particles is restricted to a region where the \vn do not change dramatically within the acceptance for the analysis and in this case the pseudorapidity dependence of  \dNdphi\ is negligible.  We consider only such analyses here, although analyses over a wide enough range in pseudorapidity for the \vn to change significantly are possible~\cite{Alver:2009id}.  The shape of a typical \dhc for trigger momenta \pttrigrange{8}{10} within pseudorapidities \etarange\ {0.5} and associated particles within \etarange\ {0.9} with momenta \ptassocrange{1.0}{2.0} in 30-40\% \Pb collisions at \sNN\ = 2.76 TeV including background \vn terms up to $n$ = 10 is shown in~\Fref{Fig:SignalPlusBackground}.  The signal is normalized by the number of \Pb collisions.  The signal is from \Fref{Fig:Signal} and the generation of the background is discussed in the Sec. III. 

In collision systems with no hydrodynamical flow where the background is due to the underlying event or soft processes uncorrelated with the reaction plane, the $J$-$B$ correlations will be independent of \dphi and therefore only add an overall constant background.  This assumption has been used for studies of \dhcs in \pp and \dAu collisions~\cite{Adams:2004wz,Adams:2005ph,Adler:2005ad,Adler:2006hi,Adler:2006sc} where this constant term is assumed to arise from the underlying event.
However, jets are correlated with the reaction plane because jets are quenched more out-of-plane, where the mean path length of medium a parton must traverse is longer, than in-plane~\cite{Afanasiev:2009aa,Adare:2012wg,Aad:2013sla}.  Therefore, when there is a background due to hydrodynamical flow and a jet signal suppressed by jet quenching, the signal and the background are both correlated with the reaction plane.  The $J$-$B$ and $B$-$J$  correlations will not be independent of \dphi in this case.  Since it is always possible to write any function as a Fourier decomposition, \Eref{Eqtn:BBCorrelations} also describes $J$-$B$ and $B$-$J$  correlations.  This leads to an overall background due to $J$-$B$, $B$-$J$, and $B$-$B$  correlations given by
\begin{equation}
 \frac{dN}{\pi d\Delta\phi} = B[ 1 + \sum_{n=1}^{\infty} 2 \tilde{v}_{n}^{\mathrm{t}} \tilde{v}_{n}^{\mathrm{a}} \cos(n\Delta\phi)],\label{Eqtn:JBBBCorrelations}
\end{equation}
\noindent where \vnefftrigger (\vneffassoc) is the pair weighted average of the \vntrigger (\vnassoc) due to jet quenching and the \vntrigger (\vnassoc) due to hydrodynamical flow.  

Generally the \vn used in background subtraction are measured separately from the correlation measurements.  The appropriate method for measuring the \vn is not obvious.  Different methods for measuring flow produce systematically different results in the same event class.  Measurements of \vn using the event plane lead to systematically higher results than measurements using correlations between multiple particles, such as a four-particle cumulant method~\cite{Voloshin:2008dg}.  The latter is less sensitive to non-flow, making it less sensitive to contamination from jets.  In principle this would make it a better measurement for the \vnassoc in \Eref{Eqtn:BBCorrelations} and \Eref{Eqtn:JBBBCorrelations}, however, these methods are also less sensitive to event-by-event fluctuations in flow and to local hot or cold spots in the medium.  Additionally, it is possible that events containing jets could have slightly different average \vn due to hydrodynamical flow than measurements of \vn in minimum bias collisions.  For these reasons it is desirable to have a method where the \vnefftrigger and \vneffassoc used for the background are determined from the same analysis as the measurement of the $J$-$J$ correlations.

Furthermore the determination of the appropriate $B$ in \Eref{Eqtn:JBBBCorrelations} is difficult and prone to assumptions about the signal which may not be true.  The most common method used is to assume a zero yield at minimum (ZYAM)~\cite{Adams:2005ph,PunchThrough,Adare:2008cqb,Ajitanand:2005jj}, or some variation of ZYAM, for instance to assume zero yield at \dphi = 1.  This assumes that there is a region in \dphi where the signal goes to zero.  The problem with this assumption is that there may be no region in azimuth where the J-J correlations go to zero.  Even in PYTHIA, at low momentum (\ptassoc $<$ 1 \GeV) there is no flat region in \dphi, indicating that there is no reliable \dphi region where the signal can be assumed to be zero.  In heavy ion collisions, where both the \ns~\cite{Aamodt:2011vg} and the \as~\cite{Adare:2012qi} peak may be modified by interactions with the medium, it is even less reliable.
The ABS method~\cite{Sickles:2009ka} uses mixed events to determine the background level.  This is an improvement on ZYAM, however, it makes the assumption that the number of $J$-$B$ and $B$-$J$  pairs are negligible compared to the number of $B$-$B$ pairs.  This assumption is valid for central collisions where the multiplicity of background particles is large, but not valid for peripheral collisions or collisions in small systems.

The data at small \dphi and large \deta are observed to be dominated by the background~\cite{Abelev:2009af,Abelev:2009ah,Agakishiev:2011st}, whatever its source, and can be fit to \Eref{Eqtn:JBBBCorrelations} to determine the \vnsq.  The background determined in this manner still assumes that the form of the background in \Eref{Eqtn:JBBBCorrelations} is correct and it is sensitive to the validity of the assumption that there is no residual signal in the large \deta and small \dphi region used for these fits.  However, it improves on ZYAM because there is no assumption that the signal goes to zero at one point and improves on both ZYAM and the ABS method because there is no assumption that the \vnsq in the background in correlation studies are equal to the $v_{n}^{\mathrm{t}}v_{n}^{\mathrm{a}}$ measured from studies of hydrodynamical flow.

\section{Method}\label{testthis}\makeatletter{}We focus on \dhcs and \jhcs with associated particle momenta \ptassocrange{1}{2} within \etarange\ {0.9} for both 0--10\% and 30--40\% central \Pb events at \sNN\  = 2.76 TeV.  We select trigger hadrons with \pttrigrange{8}{10} with $|\eta|<$ 0.5 and trigger partons with \pttrigrange{20}{40} with $|\eta|<$ 0.5.   While the ALICE detector can select trigger hadrons over a wider $\eta$ acceptance, using the same $\eta$ selection for trigger hadrons and partons simplified the simulations of the background.  
 The signal is generated from PYTHIA~\cite{Sjostrand:2006za} events using the Perugia 2011 tune~\cite{Skands:2010ak}.

 Di-hadron correlations are calculated using charged hadrons for both the trigger and associated particles.  
Jet-hadron correlations are calculated using gluons and quarks as a proxy for fully reconstructed jets.  We do not attempt a realistic simulation of jets. 
While real data could lead to fake jets, they would comprise particles correlated by flow and therefore would not have an associated \ns peak.  Instead, fake jets would change the \vnefftrigger.  Since we fit the \vnefftrigger, we would extract the correct \vnefftrigger for background subtraction using our fit. 

The background is generated assuming that each trigger and each associated particle is correlated with the reaction plane with the \vn given in \Tref{Tab:vn} up to $n$ = 10.  The available data guide the choice of \vn~\cite{CMS:2013bza,Aad:2013sla,Abelev:2012di,ALICE:2011ab}.  The exact choice of \vn does not impact whether or not the method is feasible.    Larger \vn make the background more difficult to extract, particularly for the higher-order \vn where the available data do not constrain the \vn significantly.  We therefore use upper bounds in order to test the method in a worst case scenario.  For associated particles and trigger particles for \dhcs, \vnum{2}, \vnum{3}, \vnum{4}, and \vnum{5} are chosen to approximate the values observed in the data~\cite{Abelev:2012di,ALICE:2011ab}.  The data available do not tightly constrain the higher-order \vn so we use $v_{n+2} = v_{n}/2$.  This is an approximate upper bound.  Only \vnum{2} is available for reconstructed jets~\cite{Aad:2013sla}.  We estimate that the \vn for jets is approximately the same as the \vn for high-\pT hadrons.

\begin{table}
\begin{center}
\caption{\vn values used for background calculations.  For $v_{n>5}$ we use $v_{n+2} = v_{n}/2$.}
\label{Tab:vn}
\begin{tabular}{c c | c| c |c |c }
&  &  \vnum{2} & \vnum{3} & \vnum{4} & \vnum{5} \\ \hline
\multirow{ 2}{*}{0-10\%} & $v_{n}^{a}$  & 0.041 & 0.030 & 0.0023 & 0.0011  \\ 
 & $v_{n}^{t}$  & 0.030 & 0.030 & 0.0150 & 0.0100  \\ [0.5ex]
\multirow{ 2}{*}{30-40\%} & $v_{n}^{a}$  & 0.134 & 0.047 & 0.0173 & 0.0092  \\ 
 & $v_{n}^{t}$  & 0.100 & 0.030 & 0.0150 & 0.0100  \\ 
\end{tabular}
\end{center}
\end{table}

To get the signal to background correct, 1660 PYTHIA events are simulated for each 0--10\% central \Pb event and 251 PYTHIA events for each 30--40\% central \Pb event, the number of binary nucleon-nucleon collisions in each \Pb collisions calculated by CMS~\cite{CMS:2012aa}.  
Since PYTHIA events include an underlying event, we subtract this background using the ZYAM method and use this as our known signal.
We emulate approximately $8 \times 10^{6}$ \mbox{0--10\%} and $16 \times 10^{6}$ 30--40\% central \Pb collisions.  

To simulate the background pairs, the reaction plane angle $\psi$ is chosen to be zero in detector coordinates and random trigger jets or hadrons and associated particles are thrown with a distribution described by \Eref{Eqtn:FlowFourierDecomposition} with the \vn given in \Tref{Tab:vn}.  
We estimate the number of associated particles with \ptassocrange{1}{2} with $|\eta|<$ 0.9
using charged hadron~\cite{Abelev:2012hxa,Abelev:2012wca} 
spectra measured by ALICE. 
We throw a random $\eta$ for both the associated particle and the trigger.  To emulate the approximate effect of $\eta$ dependent \vn, we apply a 1\% linear decrease in the \vn from $\eta$ = 0 to $\eta$ = 0.9, consistent with the slight $\eta$ dependence observed at midrapidity.  
The observed \vn will be largest when the $n$th-order event plane is used, however, in a typical analysis to measure the correlations due to jets, the second-order event plane is used.  
In our model we assume that all event planes for even $n$ are identical to the second-order event plane.
There is no correlation between the odd and even $n$-event planes.  
We therefore choose a random orientation for the odd $n$-event plane for each simulated event.  

The RPF method uses the reaction plane dependence to determine the background.  The finite resolution for reconstructing the reaction plane changes the shape of the background for reaction plane dependent correlation studies~\cite{Bielcikova:2003ku}.  To simulate a realistic measurement, the true reaction plane angle $\psi$ is smeared with a Gaussian with a width of 20$^\circ$ for 30--40\% central collisions and 40$^\circ$ for 0--10\% central collisions.  This reaction plane resolution is quantified in terms of
\begin{equation}
 r_{n} = < cos [n (\psi_{true}-\psi_{reco})]>.
\end{equation}
For perfect reaction plane reconstruction \Rn = 1 and for no reaction plane resolution \Rn = 0.  In our model we get \R{2} = 0.79, \R{4} = 0.38, and \R{6} = 0.11 for 30--40\% collisions and \R{2} = 0.58, \R{4}$\approx$ 0, and \R{6}$\approx$ 0 for 0--10\% collisions.  For odd $n$, \Rn = 0 because the odd and even $n$ reaction planes are not correlated.  
Data indicate that the even reaction planes are not 100\% correlated~\cite{Aad:2014fla}.  This changes the effective \vn when an analysis is done for a trigger fixed relative to the reaction plane, however, this is taken into account by using the \Rn measured relative to the reaction plane used in the analysis.

\begin{figure}
\begin{center}
\rotatebox{0}{\resizebox{8cm}{!}{
        \includegraphics{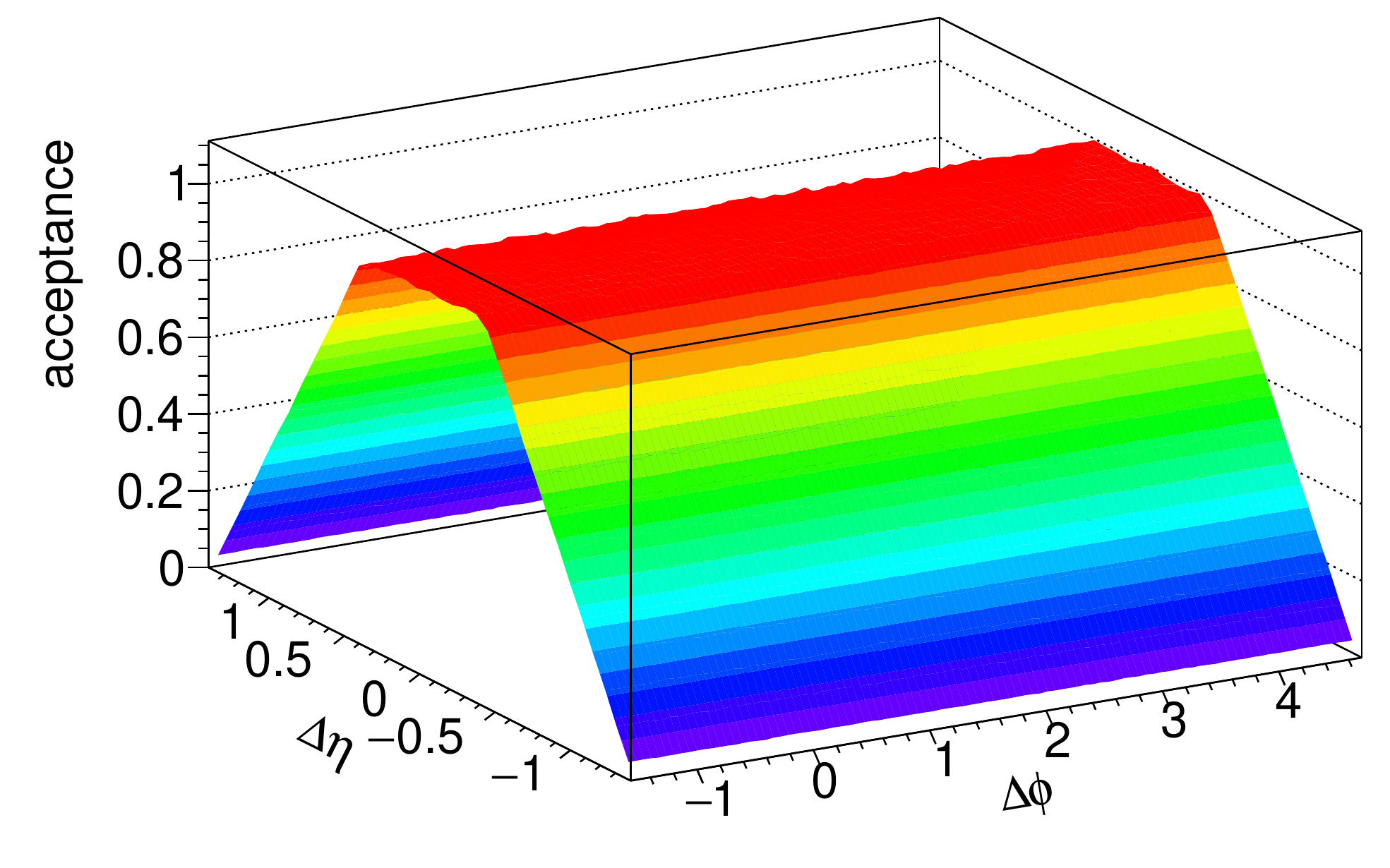}
}}\caption{Acceptance correction for a trigger with a flat distribution within $|\eta|<0.5$ and an associated particle with a flat distribution within $|\eta|<0.9$.  }\label{Fig:MixedEvents}
\end{center}
\end{figure}

The sharp cut off in the pseudorapidity $\eta$ of the particles accepted leads to a trivial structure unrelated to physics.  Pairs with \deta = $\eta^{t} - \eta^{a}$ $\approx$ 0 have nearly 100\% acceptance, however, pairs with $|\Delta\eta| \approx \eta_{max}^{t} + \eta_{max}^{a}$ have nearly 0\% acceptance.  In measurements of correlations, this is corrected with a mixed event correction which includes detector effects.  While our model has no detector effects, this trivial acceptance effect significantly modifies the simulated signal.  We therefore also apply an acceptance correction.  
If $\eta_{max}^{t} \neq \eta_{max}^{a}$, there will be a plateau between $- |\eta_{max}^{t} - \eta_{max}^{a}| $ and $ |\eta_{max}^{t} - \eta_{max}^{a}| $.  This is shown in \Fref{Fig:MixedEvents} for a trigger with a flat distribution within $|\eta|<0.5$ and an associated particle with a flat distribution within $|\eta|<0.9$. 
\section{Results}
\makeatletter{}
We first test the NSF method.  The raw signal in \Fref{Fig:SignalPlusBackground} is projected from 1.0 $<$ \deta $< $ 1.4.  This is then normalized by the \deta width of this projection in order to retain roughly the same scale as in \Fref{Fig:SignalPlusBackground}, independent of the range of the projection.  The extracted background is compared to the true background and the background extracted using a variant of the ZYAM method.  The signal is then extracted using the fit background and compared to the true signal and two variants of the ZYAM method.  

The same procedure is followed for the reaction plane dependence, testing the method with different fit ranges, for 30--40\% central collisions, and for di-hadron and \jhcs.  The same \deta range and normalizations are used for the projections.  In order to make the discussion easier to follow, we use \dhcs in 30--40\% central collisions with a fit range of \dphirange\ {$\pi$/2} as our primary reference and only vary one condition for the fit at a time.  For each sample, the true and extracted backgrounds are compared and then the true and extracted signals are compared.  The same symbols are used throughout the discussion for clarity.

The four methods used for the background subtraction are:
\begin{itemize}
 \item The ZYA1 method, a variation of ZYAM where the background is fixed at \dphi = 1 instead of at the minimum;
 \item The modified ZYA1 method, a variation of ZYA1 where the background is fixed using only data in the background-dominated region, 1.0 $<|$\deta$|<$ 1.4;
 \item The NSF method, which fits the \ns in the background-dominated region to determine the background;
 \item The RPF method, which fits the reaction-plane-dependent \ns in the background-dominated region to determine the background.
\end{itemize}
The ZYA1 method is less sensitive to statistical fluctuations than ZYAM.  The modified ZYA1 is less sensitive to the signal than ZYA1 and the background determined from this method can be directly compared to the NSF and RPF methods because they all use the same data.  The ZYA1 method requires the \vn as input.  In an analysis of data, the \vn are typically taken from other studies and the \vn have an uncertainty.  In our analysis we assume that this uncertainty is 5\%, comparable with uncertainties on \vn measured with a single method, assume that the uncertainties for the trigger and associated particles are correlated, and use the true value as the nominal value.  Since the methods for measuring \vn vary in their sensitivity to fluctuations and non-flow contributions, this likely underestimates the uncertainty on the proper \vn to use for the background in a \dhc measurement.  In addition, the nominal value of \vn used for background subtraction is not likely to be centered at the exact true \vn.  This therefore likely underestimates the uncertainties in the ZYA1 method.

We compare the true yield to the yield extracted using various methods.  The yield is given by
\begin{equation}
 Y =  \frac{dN}{N_{e}d\Delta\eta} =  \int_{a}^{b} \frac{d^2 N}{N_{e} d\Delta\eta d\Delta\phi} d\Delta\phi \label{Eq:Yields}
\end{equation}
\noindent where a = -1.05 and b = 1.05 for the \ns and a = 2.09 and b = 4.19 for the \as.  
The yields are not comparable between 0--10\% and 30--40\% central data because of the normalization by the number of events.

\subsection{The Near-Side Fit Method}
\makeatletter{}
\begin{figure}
\begin{center}
\rotatebox{0}{\resizebox{8cm}{!}{
        \includegraphics{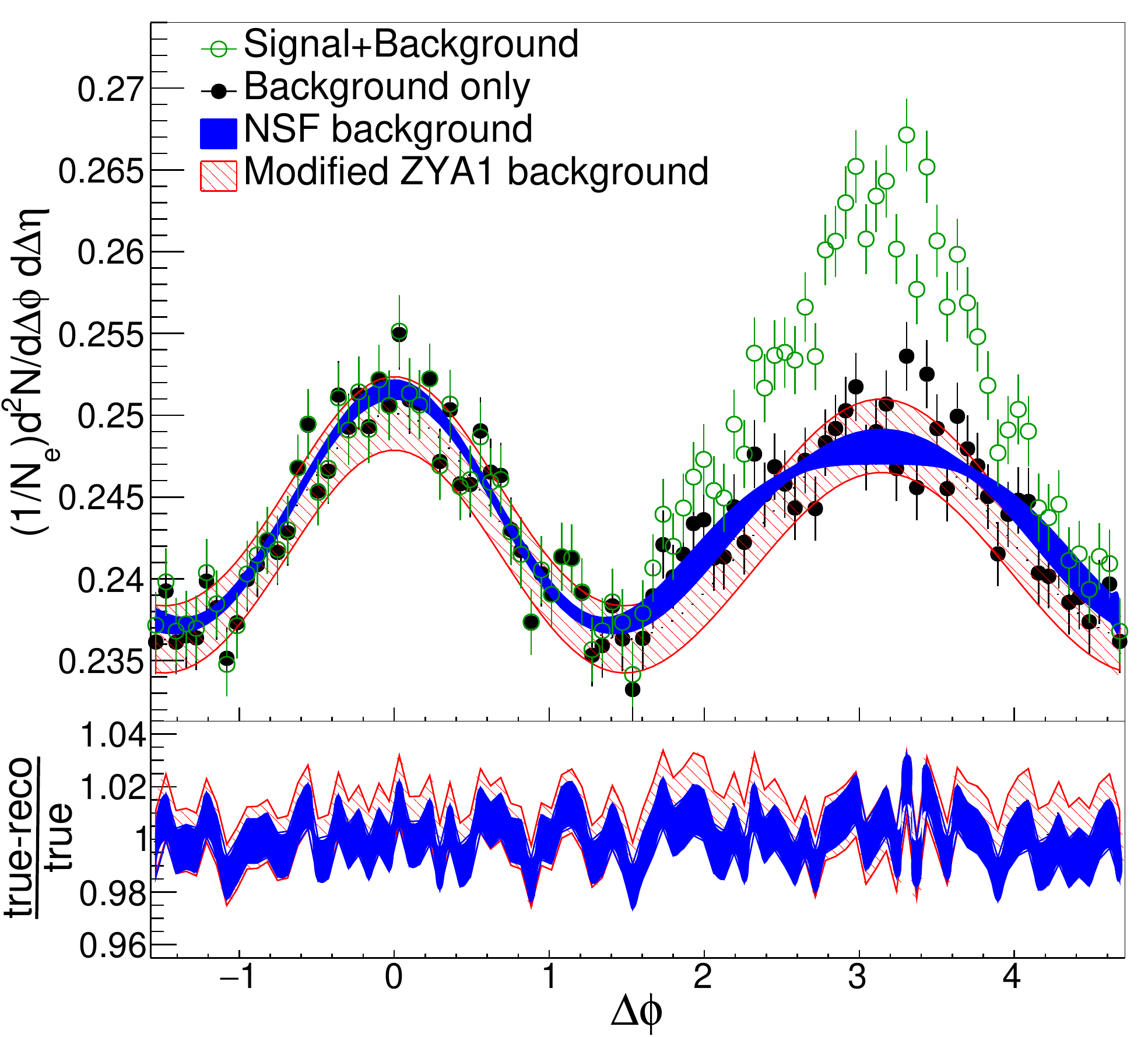}
}}
\caption{
Top:  Signal + background for \dhcs in 30--40\% \Pb collisions at \sNN\ = 2.76 TeV in the region 1.0 $<|$\deta$|<$ 1.4.  This is compared to the true background, the background from the modified ZYA1 method, and the background from the NSF method.  (See text for details.)  The fit for the NSF method is to \Eref{Eqtn:JBBBCorrelations} to order n = 4 from \dphirange\ {$\pi$/2} and has \chisq{63.6}{45}.  Bottom:  Ratios of the background from the NSF and ZYA1 methods to the true background.  }\label{Fig:FitAloneSignalPlusBackground}
\vspace{-0.20cm}
\end{center}
\end{figure}

The raw signal in the region 1.0 $<|$\deta$|<$ 1.4 is fit to \Eref{Eqtn:JBBBCorrelations} to order n = 3 from \dphirange\ {$\pi$/2}.  \Fref{Fig:FitAloneSignalPlusBackground} shows the true background, the signal plus background, 
 and the background from this fit
for \dhcs in 30--40\% \Pb collisions at \sNN\ = 2.76 TeV.  
\Fref{Fig:FitAloneSignalPlusBackground} also shows the background extracted in this region using the modified ZYA1 method.

\begin{figure}
\begin{center}
\rotatebox{0}{\resizebox{8cm}{!}{
        \includegraphics{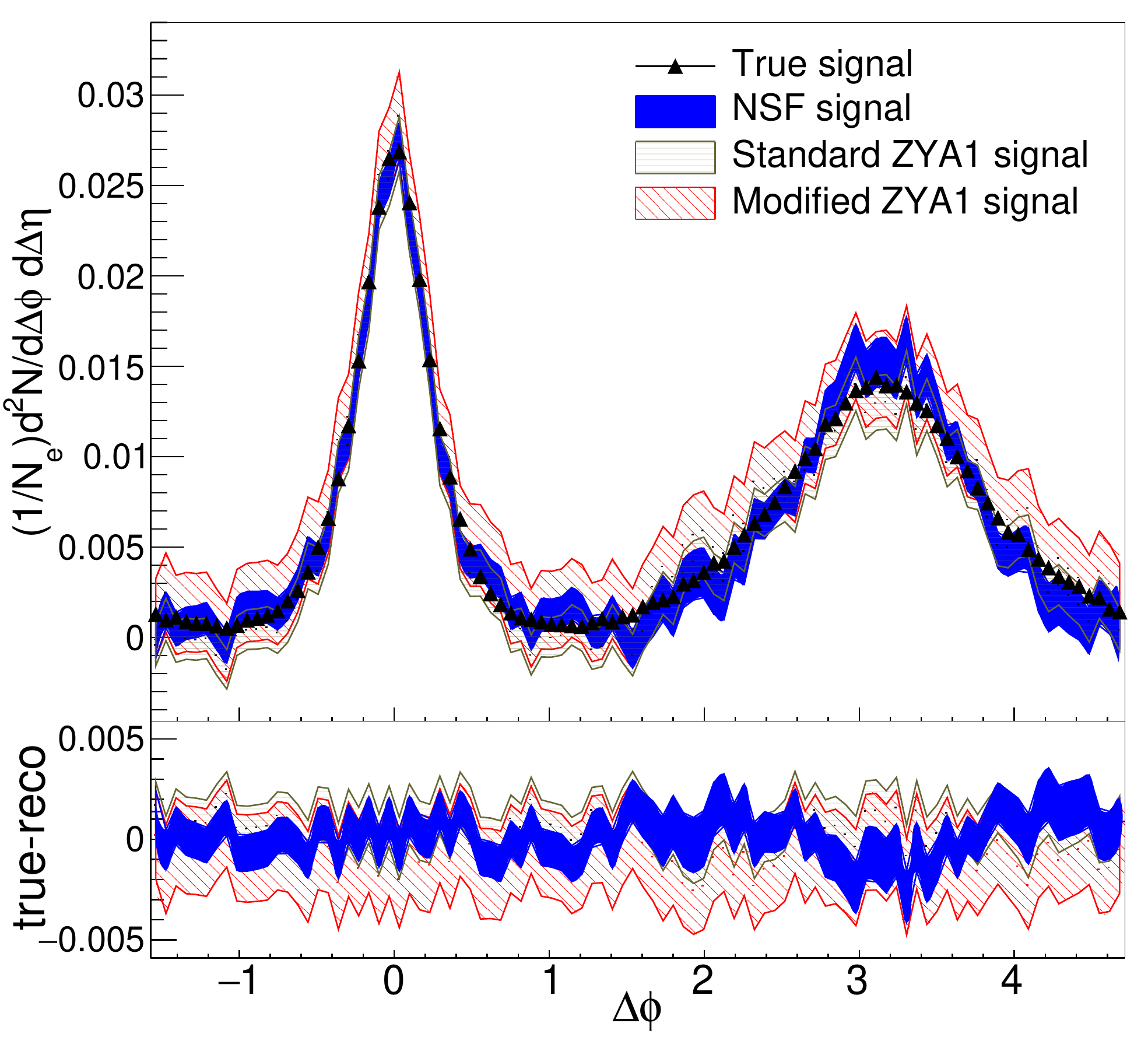}
}}
\caption{
Top:  The true signal for \dhcs in 30--40\% \Pb collisions at \sNN\ = 2.76 TeV.  This is compared to the signal extracted using the background from the ZYA1 method, the modified ZYA1 method, and the background extracted from the NSF method for \dphirange\ {$\pi$/2} using the fit shown in \Fref{Fig:FitAloneSignalPlusBackground}.  (See text for details.)  Bottom:  Differences between the true signal and the signal extracted using the background from the ZYA1 method, modified ZYA1 method, and the background from the RPF method.  }\label{Fig:FitAloneSignalExtracted}
\end{center}
\end{figure}

The signal extracted using the NSF background, the modified ZYA1 background, and the standard ZYA1 background are compared in \Fref{Fig:FitAloneSignalExtracted} to the true signal.  Only statistical uncertainties on the background are shown.  \Fref{Fig:FitAloneSignalPlusBackground} and \Fref{Fig:FitAloneSignalExtracted} show that the NSF, ZYA1, and modified ZYA1 methods describe the background well in this model and that they have comparable uncertainties.  The nominal value of the signal extracted using the fit is not centered on the true value like the nominal values for the ZYA1 and modified ZYA1 methods, however, as discussed above, our implementation of these methods may slightly underestimate the true uncertainties on the \vn and therefore underestimate the shape distortions from the ZYA1 method.  Additionally, in a heavy ion collision, the \as could be modified significantly and become much broader.  In this scenario, the NSF method would be less sensitive to residual signal in the \as than the ZYA1 method.

\begin{figure}
\begin{center}
\rotatebox{0}{\resizebox{8cm}{!}{
   \includegraphics{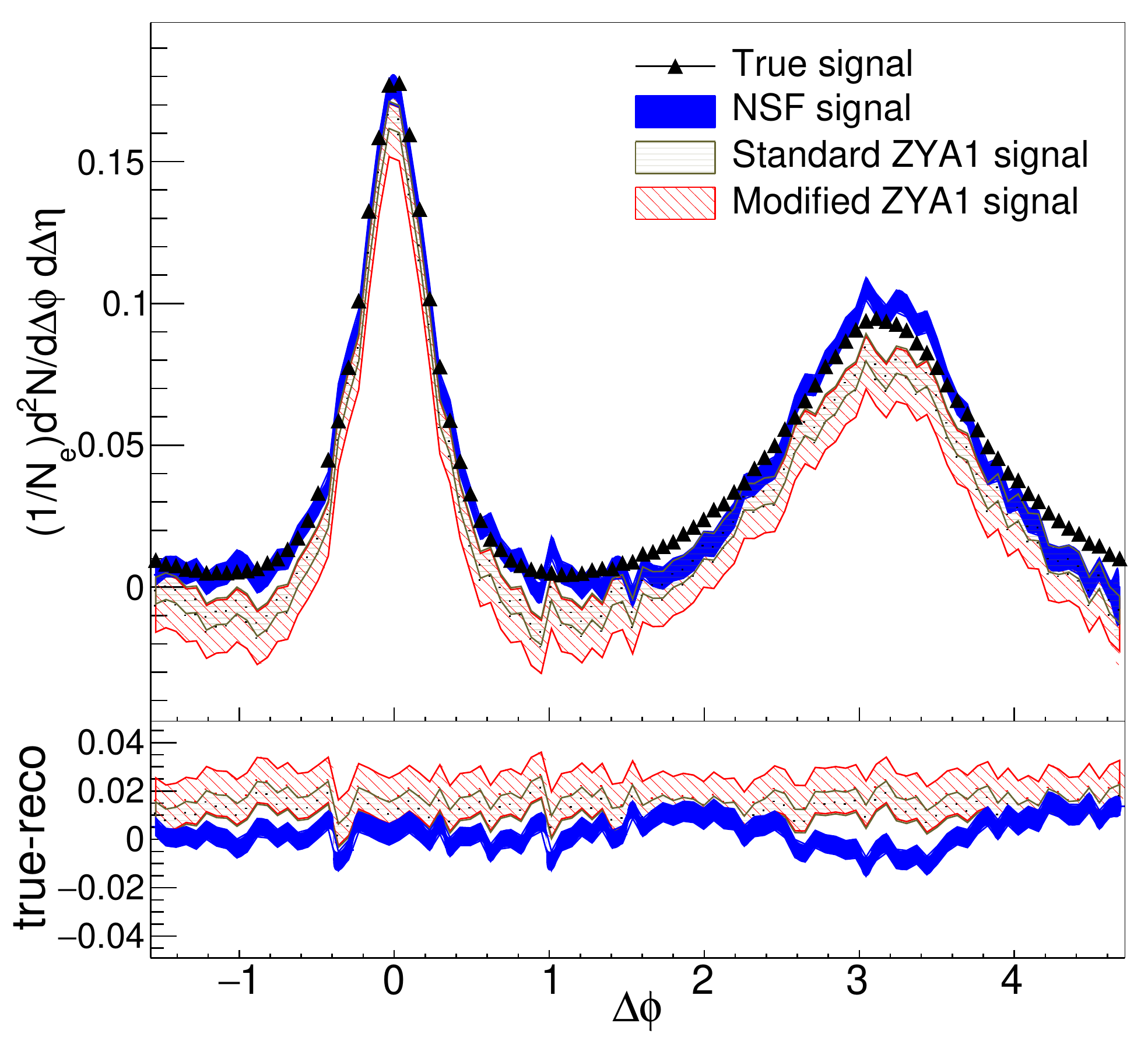}
 }}
\caption{ Top:  The true signal for \dhcs in 0--10\% \Pb collisions at \sNN\ = 2.76 TeV.  This is compared to the signal extracted using the background from the ZYA1 method, the modified ZYA1 method, and the background extracted from the NSF method for \dphirange\ {$\pi$/2}.
Bottom:  Differences between the true signal and the signal extracted using the background from the ZYA1 method, modified ZYA1 method, and the background from the RPF method.  }\label{Fig:FitAloneSignalExtracted_hhCentral}
\end{center}
\end{figure}

\begin{figure}
\begin{center}
\rotatebox{0}{\resizebox{8cm}{!}{
   \includegraphics{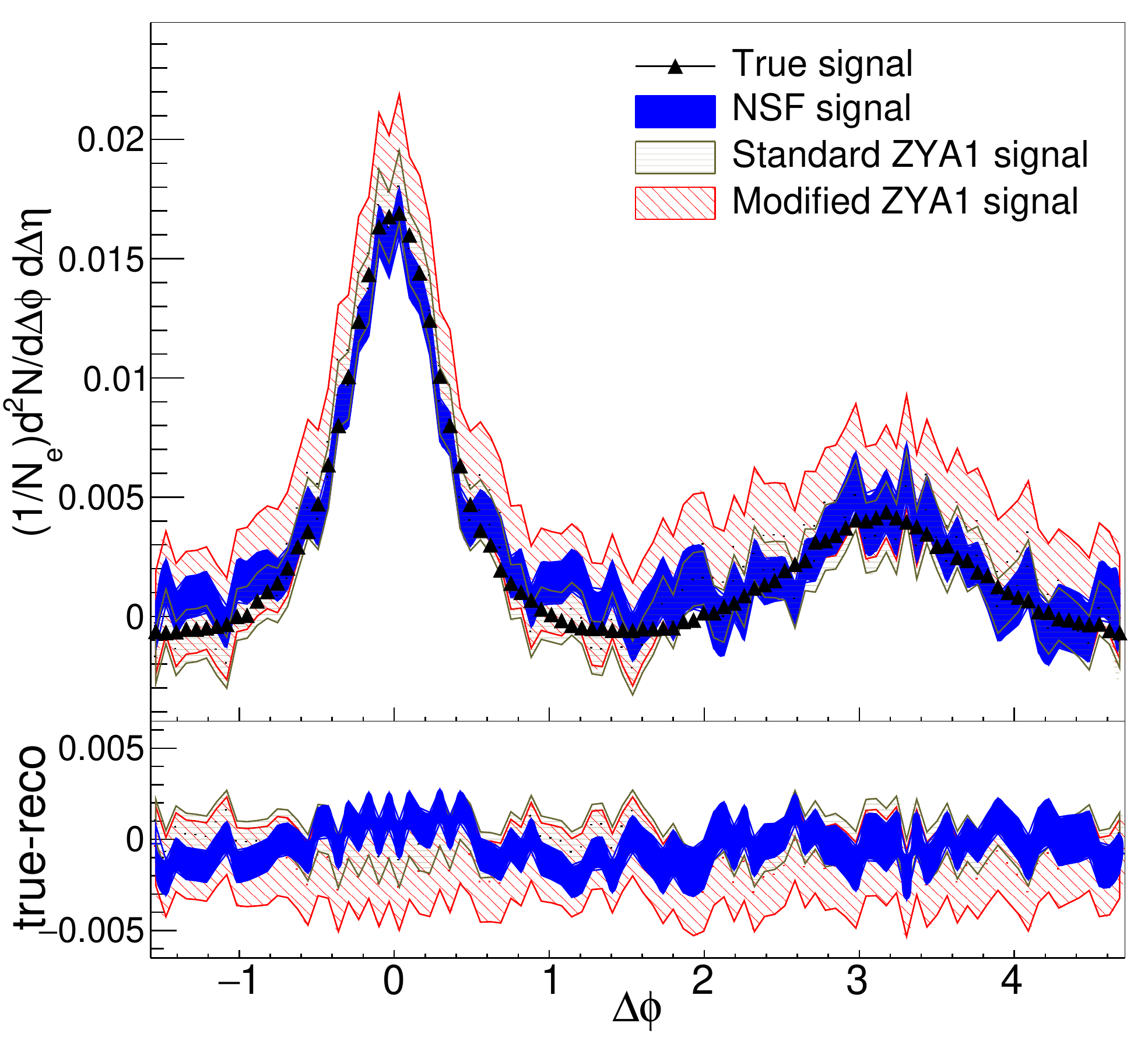}
 }}
\caption{ Top:  The true signal for \jhcs in 30--40\% \Pb collisions at \sNN\ = 2.76 TeV.  This is compared to the signal extracted using the background from the ZYA1 method, the modified ZYA1 method, and the background extracted from the NSF method for \dphirange\ {$\pi$/2}.
Bottom:  Differences between the true signal and the signal extracted using the background from the ZYA1 method, modified ZYA1 method, and the background from the RPF method.  }\label{Fig:FitAloneSignalExtracted_jethMidPeripheral}
\end{center}
\end{figure}

 \Fref{Fig:FitAloneSignalExtracted_hhCentral} and \Fref{Fig:FitAloneSignalExtracted_jethMidPeripheral} compare the signal extracted using the ZYA1 method, modified ZYA1 method, and NSF method  for \dphirange\ {$\pi$/2} for \dhcs in \mbox{0--10\%} central \Pb collisions and \jhcs in 30--40\% central \Pb collisions.  The NSF method describes the background better for 0--10\% central collisions than for 30--40\% central collisions.  In central collisions the \vneff are smaller and the background is larger so it is possible to determine the background with higher precision. The yields are extracted for all these cases using \Eref{Eq:Yields} and are summarized in the \Tref{Tab:YieldsNoRxnPlane}. The NSF method provides a more precise measurement of the yield than ZYA1 in all cases.

\begin{table}
\begin{center}
\caption{
Yields $Y$ as defined in \Eref{Eq:Yields} scaled by 10$^{-3}$ from 
\Fref{Fig:FitAloneSignalExtracted}, \Fref{Fig:FitAloneSignalExtracted_hhCentral}, and \Fref{Fig:FitAloneSignalExtracted_jethMidPeripheral}.  The first uncertainty on the true yield is the statistical uncertainty and the second is the uncertainty due to the uncertainty on the subtraction of the background from the underlying event.  The first uncertainty for the ZYA1 method is the statistical uncertainty, the second uncertainty is the uncertainty on the background level, and the third uncertainty is the uncertainty due to \vnumeffassoc{2} and \vnumefftrigger{2}.  ZYA1 uncertainties are propagated assuming 100\% correlation between \vnumeffassoc{2} and \vnumefftrigger{2} and no correlation between the uncertainty on the level of the background and the uncertainty on the \vn.  Errors due to higher order \vn are not considered but are approximately 10\% of the uncertainties due to \vnumeffassoc{2} and \vnumefftrigger{2}.
}
\label{Tab:YieldsNoRxnPlane}
\resizebox{\linewidth}{!}{ 
\begin{tabular}{c |c | c | c }
\hline
\hline
\multirow{2}{*}{ Sample }&   & \multicolumn{2}{c}{Yield ($Y\times 10^{-3}$)} \\ 
  &  & \ns  & \as \\

\hline
 & True                            & 17.1 $\pm$ 0.1 $\pm$ 0.2 & 19.9 $\pm$ 0.1 $\pm$ 0.2\\
30--40\% & Mod. ZYA1 & 18.9 $\pm$ 4.2 $\pm$ 1.2 & 21.9 $\pm$ 4.2 $\pm$ 1.2\\ 
h-h & Std. ZYA1 & 15.7 $\pm$ 1.6 $\pm$ 1.2 & 18.7 $\pm$ 1.6 $\pm$ 1.2\\ 
 & NSF &  17.14 $\pm$  1.1   &  20.14 $\pm$ 1.11 \\

\hline
& True                              & 114.4 $\pm$ 0.2 $\pm$ 0.8  & 132.8 $\pm$ 0.2 $\pm$ 0.8\\
0--10\% & Mod. ZYA1 & 75.5 $\pm$ 18.3 $\pm$ 0.9  &  95.7 $\pm$ 18.3 $\pm$ 0.9\\ 
h-h  & Std. ZYA1 & 86.7 $\pm$ 7.0 $\pm$ 0.9 & 106.9 $\pm$ 7.0 $\pm$ 0.9\\ 
& NSF  &  111.63 $\pm$  3.01  &  131.82 $\pm$ 3.01 \\

 \hline
& True                             &  13.19 $\pm$ 0.04 $\pm$ 0.17  &  4.96 $\pm$ 0.04 $\pm$ 0.17\\ 
30--40\% & Mod. ZYA1 & 16.2 $\pm$ 4.2 $\pm$ 1.2 & 8.2 $\pm$ 4.2 $\pm$ 1.2\\ 
jet-h & Std. ZYA1 & 13.2 $\pm$ 1.6 $\pm$ 1.2 & 5.2 $\pm$ 1.6 $\pm$ 1.2\\
& NSF  &   13.13 $\pm$  0.77 &  5.13 $\pm$ 0.78\\

\hline
\hline
\end{tabular}
}
\end{center}
\end{table}

\begin{figure}
\begin{center}
  \vspace{0.70cm}
\rotatebox{0}{\resizebox{8cm}{!}{
        \includegraphics{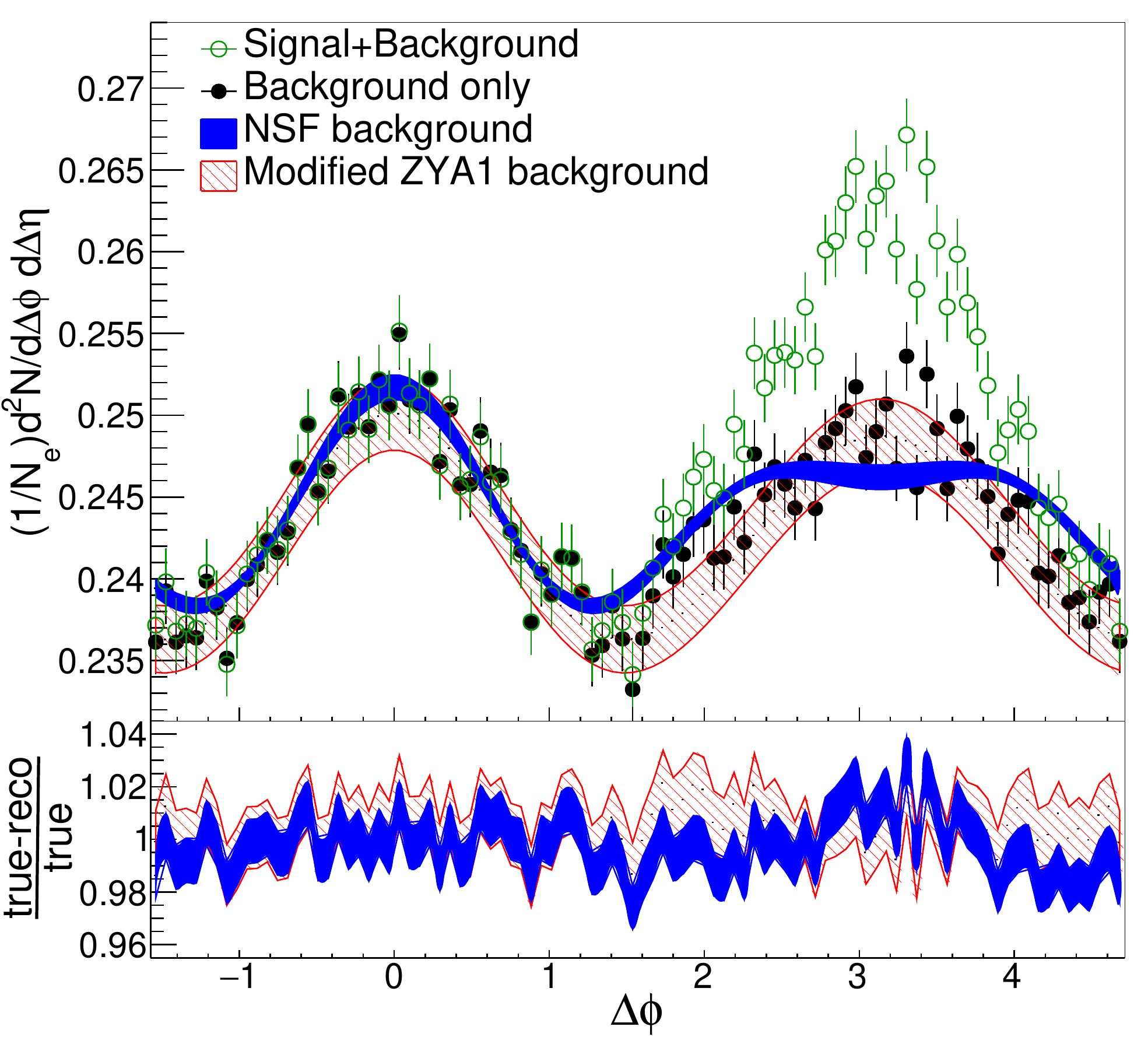}
}}
\caption{
Top:  Signal+background for \dhcs in 30--40\% \Pb collisions at \sNN\ = 2.76 TeV in the region 1.0 $<|$\deta$|<$ 1.4.  This is compared to the true background, the background from the modified ZYA1 method, and the background from the NSF method.  (See text for details.)  The fit for the NSF method is to \Eref{Eqtn:JBBBCorrelations} to order n = 4 from \dphirange\ {1.25} and has \chisq{50.8}{35}.  Bottom:  Ratios of the background from the NSF and ZYA1 methods to the true background.  }\label{FitAloneBackgroundNarrowdPhi}
\vspace{-0.40cm}
\end{center}
\end{figure}

However, we noticed that the fit was sensitive to the fit range.  \Fref{FitAloneBackgroundNarrowdPhi} shows the true background, the signal+background, the background using the ZYA1 method, and the background extracted using the NSF method in the range \dphirange\ {1.25}.  This fit clearly fails to describe the \vn background, even though the fit converged and the fit quality is comparable to the fit in the range \dphirange\ {$\pi$/2}.  This is because the fit needs to be able to distinguish between \vnum{2} and \vnum{3}.  At \dphi = $\pi$/3, the \vnum{3} term starts increasing while the \vnum{2} term is still decreasing.  Either the fit needs to cover enough range in \dphi to distinguish between these two terms or the data need to have enough statistics that it is possible to discern the relative weights of the \vnum{2} and \vnum{3} terms from the width of the \ns peak.  The fit in \Fref{FitAloneBackgroundNarrowdPhi} shows that realistic statistics do not provide data with the precision required for the latter.  If a fit to the \ns at large \deta were used for analyzing data, it could lead to subtracting the wrong background and potentially extracting a signal with a distorted signal on the \as.  We therefore explore using the reaction plane dependence of the raw signal, which uses more information and therefore is likely to have lower uncertainties and be more stable.

\subsection{The Reaction Plane Fit Method}
\makeatletter{}When a trigger hadron or parton is restricted relative to the reaction plane, the level of the background and the effective \vntrigger are  affected.  The derivation of the appropriate reaction plane dependent forms are discussed in Ref.~\cite{Bielcikova:2003ku}.
When the trigger is restricted to a range of angles relative to the reconstructed reaction plane, the effective even \vntrigger are given by
\begin{widetext}
\begin{equation}
 \tilde{v}_n^{R,t} = \frac{v_n+\cos(n\phi_S)\frac{\sin(nc)}{nc} R_n + \sum_{k=2,4,6...} (v_{k+n}+v_{|k-n|}) \cos(k\phi_S) \frac{\sin(kc)}{kc} R_n }{1+ \sum_{k=2,4,6...} 2 v_k \cos(k \phi_S) \frac{\sin(kc)}{kc} R_n}\label{Eq:vnEff}
 \end{equation}
\end{widetext}

 \noindent and the effective background level is given by 
 \begin{equation}
    \tilde{\beta^{R}} = 1 + \sum_{k=2,4,6...} 2 v_k \cos(k\phi_S) \frac{\sin(kc)}{kc} R_n \label{Eq:BEff}
 \end{equation}

\noindent where $\phi_S$ is the center of range and $2c$ is the width of the range~\cite{Bielcikova:2003ku}.  The background is then given by:

\begin{equation}
 \frac{dN}{\pi d\Delta\phi} = \tilde{\beta^{R}} ( 1 + \sum_{n=1}^{\infty} 2 \tilde{v}_{n}^{R,\mathrm{t}} \tilde{v}_{n}^{\mathrm{a}} \cos(n\Delta\phi)).\label{Eqtn:JBBBCorrelationsRxnPlane}
\end{equation}

\begin{figure}
\begin{center}
\rotatebox{0}{\resizebox{8cm}{!}{
        \includegraphics{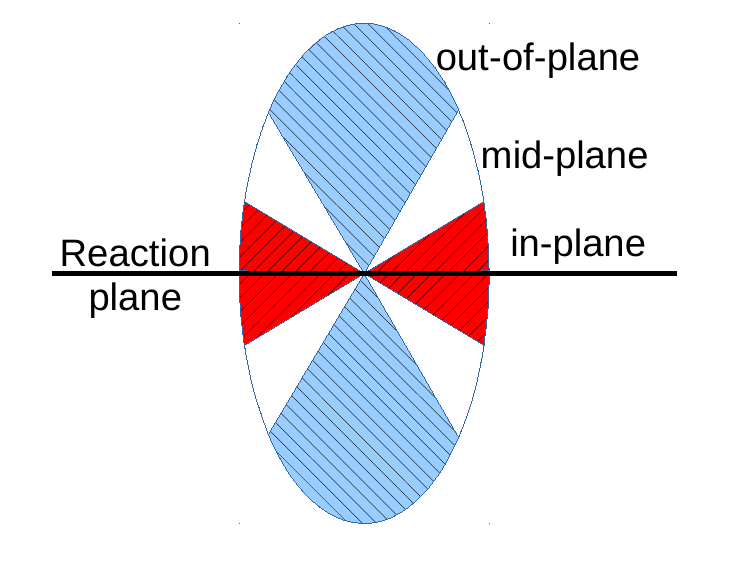}
}}
\caption{Schematic diagram showing the reaction plane angles used in the analysis.  }\label{Fig:RxnPlaneFig}
\end{center}
\end{figure}

\noindent  Since the reaction planes for odd $n$ are uncorrelated with the n = 2 reaction plane, all odd $n$ terms have \vneffR = \vneff when the n = 2 reaction plane is used for an analysis.  Here we consider simultaneous measurements of \dNeventdphideta with the trigger restricted to four different regions:
\begin{itemize}
 \item All;
 \item \Inplane: $\phi_S$ = 0, c=$\pi/6$;
 \item \Midplane: $\phi_S$ = $\pi/4$ and $\phi_S$ = $3\pi/4$, c=$\pi/12$; 
 \item \Outplane: $\phi_S$ = $\pi/2$, c=$\pi/6$.
\end{itemize}
\noindent  Note that the \midplane range is actually split into four symmetric regions.  These regions are shown schematically in \Fref{Fig:RxnPlaneFig}.  The information on the reaction plane dependence of the raw correlations can reduce the uncertainty on the background.  This can be understood by considering what additional information the reaction plane dependence provides.  The level of the \inplane correlation is increased by \vnumtrigger{2} and the \vnumeffR{2} term is increased, as shown in \Eref{Eq:vnEff} and \Eref{Eq:BEff}.  This allows \vnumtrigger{2} to be determined with high precision from these correlations.  In contrast, the background level of the \midplane correlation is insensitive to \vnumtrigger{2} and the modulation of the correlation by \vnumeffR{2} is approximately equal to \vnumtrigger{2}.  Since the \inplane and \outplane correlations strongly constrain \vnumtrigger{2}, the \midplane correlation can be used to constrain the higher-order \vn.  
The normalization of the correlations per event allows the constant $\beta_R$ to be the same for all reaction plane orientations.

\subsubsection{\Dhcs in 30-40\% central \Pb collisions with the RPF method over \dphirange{$\pi$/2}}

\begin{figure*}
\begin{center}

\rotatebox{0}{\resizebox{17cm}{!}{
        \includegraphics{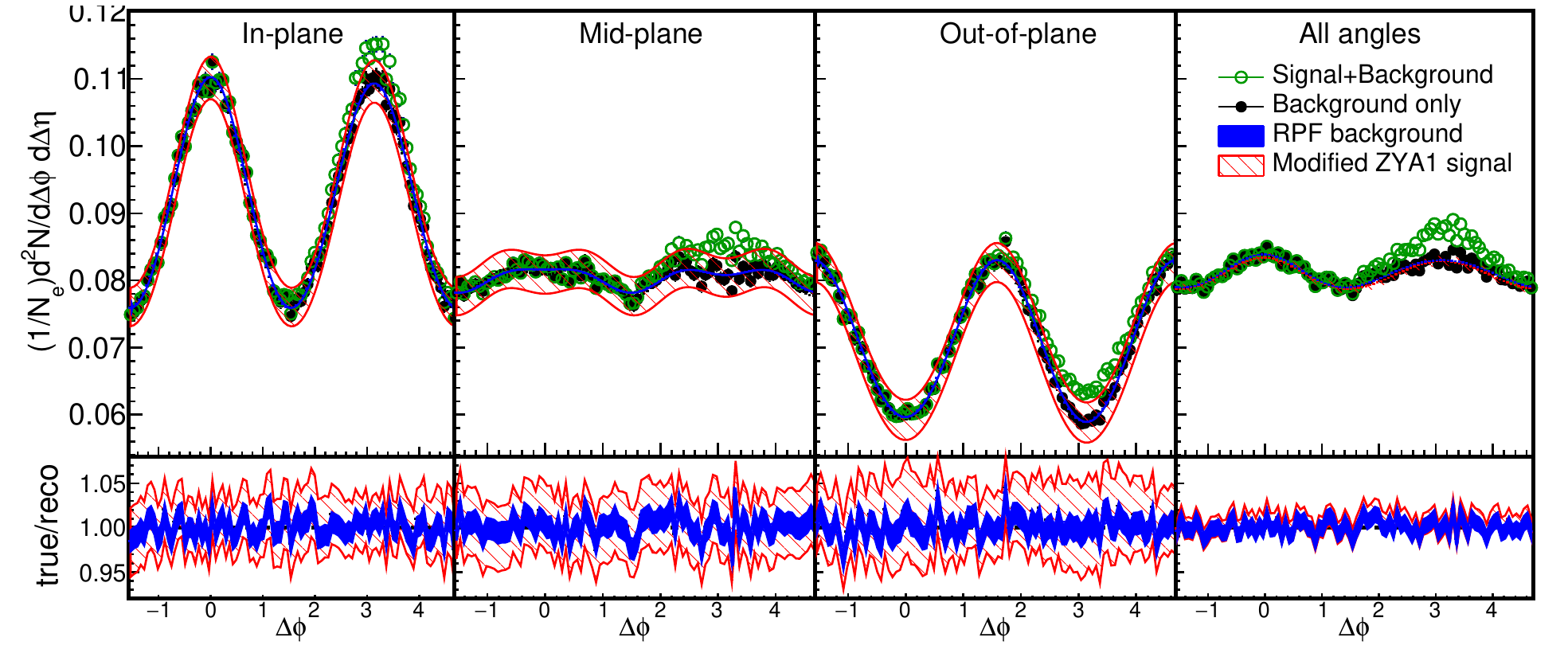}
}}\caption{
{
Top:  Signal+background for \dhcs in 30--40\% \Pb collisions at \sNN\ = 2.76 TeV in the region 1.0 $<|$\deta$|<$ 1.4 for \inplane, \midplane, and \outplane triggers and for all triggers combined.  This is compared to the true background, the background from the modified ZYA1 method, and the background from the RPF method.  (See text for details.)  The data for all angles relative to the reaction plane have been scaled by 1/3 in the top panel in order to fit on the same scale.  The fit for the RPF method is to \Eref{Eqtn:JBBBCorrelationsRxnPlane} to order n = 4 from \dphirange\ {$\pi$/2} and has \chisq{176}{138}.  Bottom:  Ratios of the background from the RPF and ZYA1 methods to the true background.  }
}\label{Fig:PiecewiseFitBackgroundExtractionhhMidPeripheral}
\end{center}
\end{figure*}

\begin{figure*}
\begin{center}
\rotatebox{0}{\resizebox{17cm}{!}{
        \includegraphics{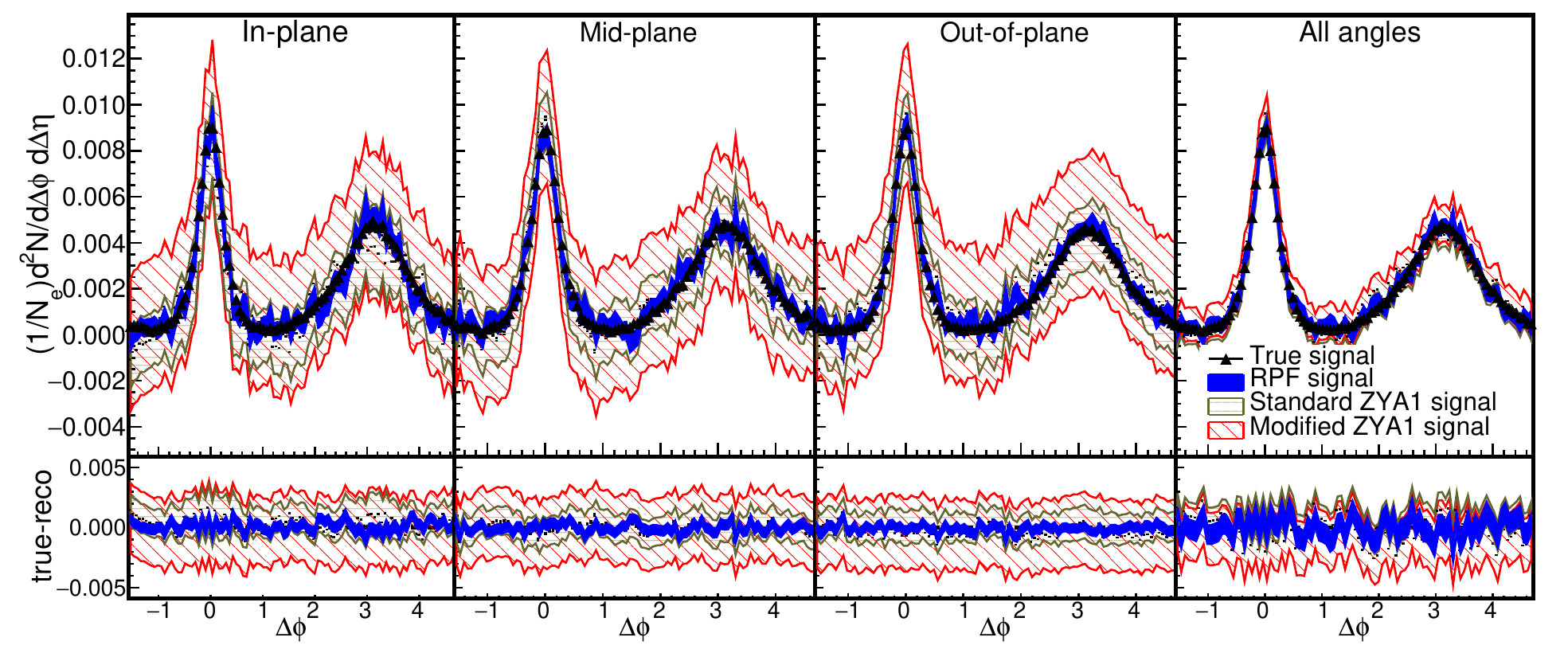}
}}
\caption{
{
Top:  The true signal for \dhcs in 30--40\% \Pb collisions at \sNN\ = 2.76 TeV for \inplane, \midplane, and \outplane triggers and for all triggers combined.  This is compared to the signal extracted using the background from the ZYA1 method, the modified ZYA1 method, and the background extracted from the RPF method for \dphirange\ {$\pi$/2} using the fit shown in \Fref{Fig:PiecewiseFitBackgroundExtractionhhMidPeripheral}.  (See text for details.)  The data for all angles relative to the reaction plane have been scaled by 1/3 in the top panel in order to fit on the same scale. Bottom:  Differences between the true signal and the signal extracted using the background from the ZYA1 method, modified ZYA1 method, and the background from the RPF method. }
}\label{Fig:PiecewiseFitTrueVsRecoSignalhhMidPeripheral}
\end{center}
\end{figure*}

\Fref{Fig:PiecewiseFitBackgroundExtractionhhMidPeripheral} shows the signal+background in the region 1.0 $<|$\deta$|<$ 1.4 for \dhcs in 30--40\% \Pb collisions at \sNN\ = 2.76 TeV for \inplane, \midplane, \outplane, and \allplane.  We fit the signal on the \ns to all reaction plane orientations simultaneously, restricting the fits for each reaction plane orientation to \dphirange\ {$\pi$/2}.  The \Rn are fixed at the values in our model.  In an experimental analysis, the \Rn can be measured.  We varied the order of \vn used in the fit until we used the fewest parameters necessary to get a reliable fit.  We found that the fit worked best to n = 4, corresponding to a total of six parameters, $B$, 
\vnumefftrigger{2}, \vnumeffassoc{2}, \vnumeffsq{3}, \vnumefftrigger{4}, and \vnumeffassoc{4}.
The extracted parameters were all within error of the parameters used in the simulation.  The RPF method is compared to the modified ZYA1 method.  Note that for the ZYA1 method a different {\color{red}\vneff} must be used for each reaction plane orientation.  Again we use the nominal values of \vn thrown using our model and assume 5\% uncertainties on the \vntrigger.  Only statistical uncertainties are shown for each background method.  \Fref{Fig:PiecewiseFitTrueVsRecoSignalhhMidPeripheral} compares the signal extracted using the various background methods to the true background.  The yields are given in \Tref{Tab:Yields}.  The RPF method leads to a much more accurate determination of the signal shape and the yield than the ZYA1 or modified ZYA1 methods for correlations with a trigger \inplane and \outplane.  This is because even with the optimistic 5\% uncertainty on the \vn, the amplification of the \vnumeff{2} for these correlations makes the accurate determination of the background difficult.

\subsubsection{\Dhcs in 30-40\% central \Pb collisions with the RPF method over \dphirange\ {1}}

\begin{figure*}
\begin{center}
\rotatebox{0}{\resizebox{17cm}{!}{
        \includegraphics{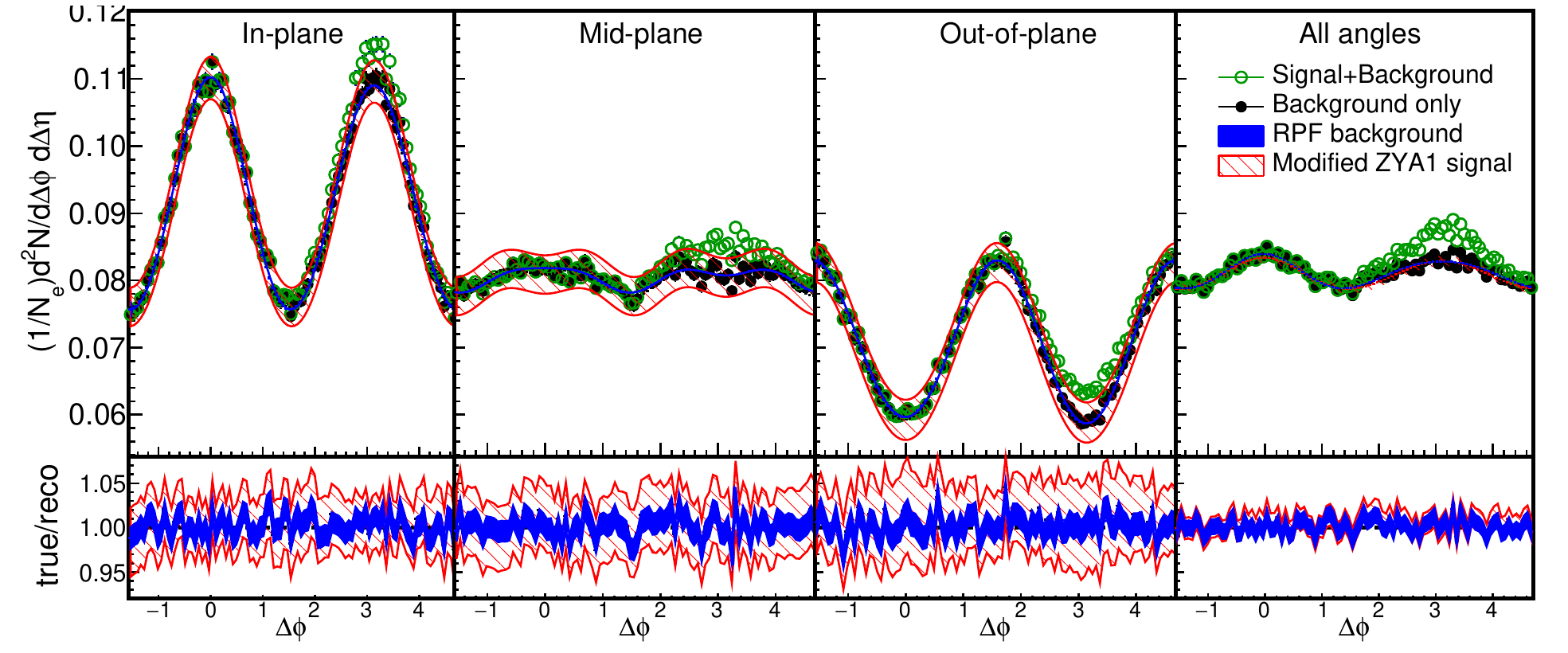}
}}\caption{
{
Top:  Signal+background for \dhcs in 30--40\% \Pb collisions at \sNN\ = 2.76 TeV in the region 1.0 $<|$\deta$|<$ 1.4 for \inplane, \midplane, and \outplane triggers and for all triggers combined.  This is compared to the true background, the background from the modified ZYA1 method, and the background from the RPF method.  (See text for details.)  The data for all angles relative to the reaction plane have been scaled by 1/3 in the top panel in order to fit on the same scale.  The fit for the RPF method is to \Eref{Eqtn:JBBBCorrelationsRxnPlane} to order n = 4 from \dphirange\ {1} and has \chisq{128}{90}.  Bottom:  Ratios of the background from the RPF and ZYA1 methods to the true background.  }
}\label{Fig:PiecewiseFitBackgroundExtractionhhMidPeripheralNarrowdPhi}
\end{center}
\end{figure*}

\begin{figure*}
\begin{center}
\rotatebox{0}{\resizebox{17cm}{!}{
        \includegraphics{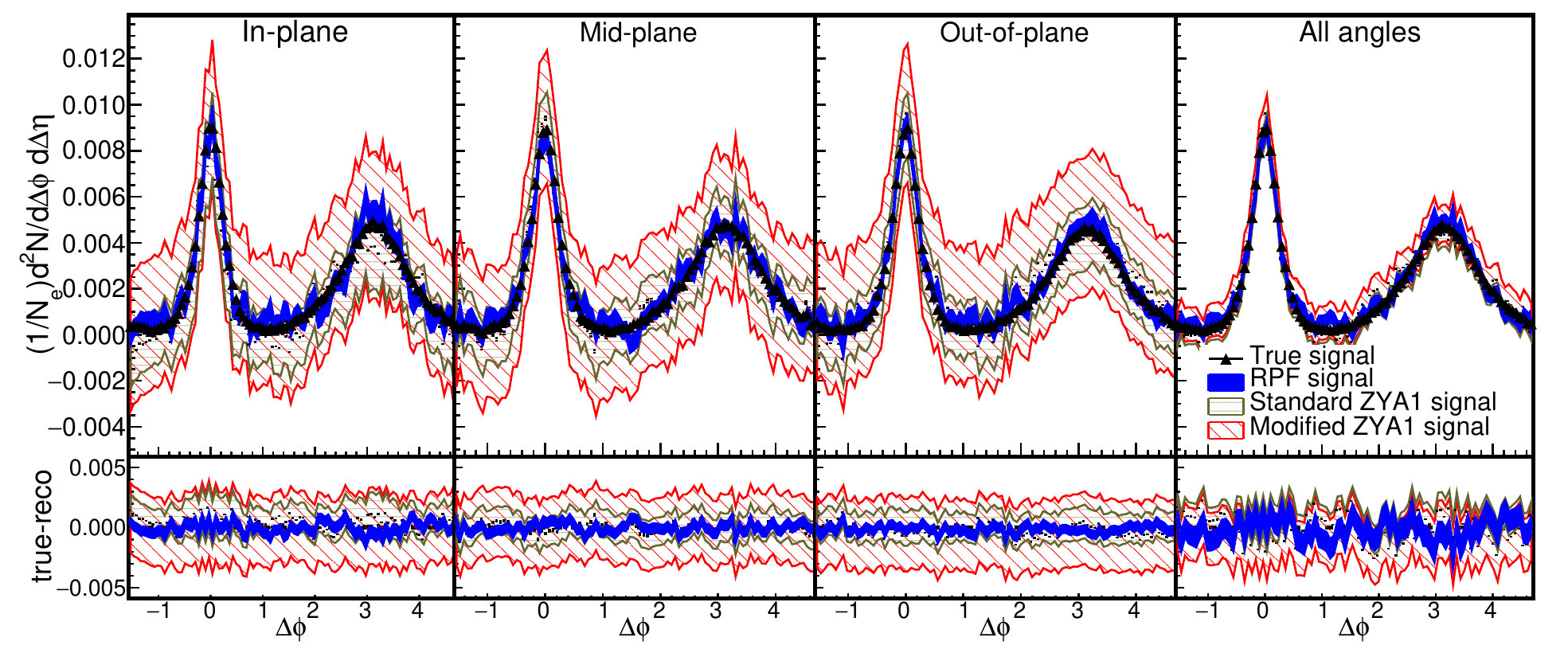}
}}
\caption{ 
Top:  The true signal for \dhcs in 30--40\% \Pb collisions at \sNN\ = 2.76 TeV for \inplane, \midplane, and \outplane triggers and for all triggers combined.  This is compared to the signal extracted using the background from the ZYA1 method, the modified ZYA1 method, and the background extracted from the RPF method for \dphirange\ {1} using the fit shown in \Fref{Fig:PiecewiseFitBackgroundExtractionhhMidPeripheralNarrowdPhi}.  (See text for details.)  The data for all angles relative to the reaction plane have been scaled by 1/3 in the top panel in order to fit on the same scale.  Bottom:  Differences between the true signal and the signal extracted using the background from the ZYA1 method, modified ZYA1 method, and the background from the RPF method.  }
\label{Fig:PiecewiseFitTrueVsRecoSignalhhMidPeripheralNarrowdPhi}
\end{center}
\end{figure*}

One of the benefits of the simultaneous fit of the background to the different reaction plane orientations on the \ns is that it may allow a fit over a narrower \dphi range, which would reduce the impact of any signal from the \as from a modified \as.  We therefore fit the signal over \dphirange\ {1}, which should be less sensitive to residual signal than a fit to \dphirange\ {$\pi$/2}.  This is a narrower range than that shown in \Fref{FitAloneBackgroundNarrowdPhi} because by varying the fit range we found that the simultaneous fit converged even when a narrower range in \dphi was used.  The background extracted with this fit is compared in \Fref{Fig:PiecewiseFitBackgroundExtractionhhMidPeripheralNarrowdPhi} to the true background and the background extracted using the modified ZYA1 method.  The signal extracted using this fit is shown in \Fref{Fig:PiecewiseFitTrueVsRecoSignalhhMidPeripheralNarrowdPhi}  and the yields are given in \Tref{Tab:Yields}.  While the signal using the fit over \dphirange\ {$\pi$/2} shown in \Fref{Fig:PiecewiseFitTrueVsRecoSignalhhMidPeripheral} has slightly smaller uncertainties, the signal using the fit over \dphirange\ {1} shown in \Fref{Fig:PiecewiseFitTrueVsRecoSignalhhMidPeripheralNarrowdPhi} is comparable.  Since the width of the \as peak varies with both \pttrig and \ptassoc and even the \ns width is observed to be modified in heavy ion collisions~\cite{Agakishiev:2011st}, a narrower fit range is better.  With a range covering \dphirange\ {1}, even an \as peak with a width of \dphi = 1 would have less than 5\% of its amplitude in the fit region.

\subsubsection{\Dhcs in 0-10\% central \Pb collisions with the RPF method over \dphirange\ {$\pi$/2}}

Central collisions are often considered the most interesting because the medium reaches higher energy densities and hottest temperatures.  Naively one might assume that this method cannot be applied to central collisions because the reaction plane is not known to better than approximately 40$^{\circ}$ and the reaction plane bins described above are 30$^{\circ}$ wide.  \Eref{Eq:vnEff} and \Eref{Eq:BEff} show that the \Beff and \vneffR differ less between \inplane, \midplane, and \outplane when the \Rn are small than when the \Rn are large.  This is because a large fraction of the trigger particles reconstructed \inplane will actually come from \midplane and some will even come from \outplane.  Still, there is a difference between the 
correlations for trigger particles reconstructed with different reaction plane angles and this provides some information.

\begin{figure*}
\begin{center}
\rotatebox{0}{\resizebox{17cm}{!}{
        \includegraphics{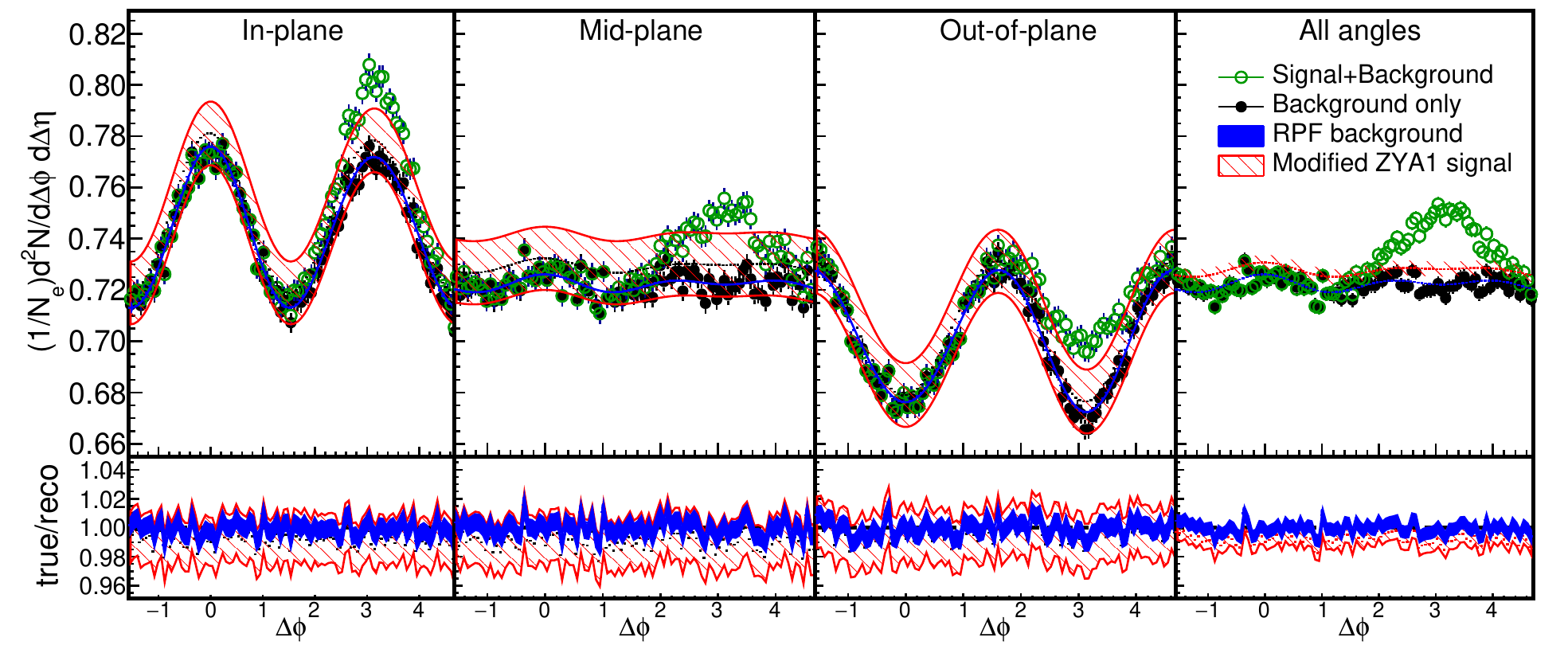}
}}\caption{
Top:  Signal+background for \dhcs in 0--10\% \Pb collisions at \sNN\ = 2.76 TeV in the region 1.0 $<|$\deta$|<$ 1.4 for \inplane, \midplane, and \outplane triggers and for all triggers combined.  This is compared to the true background, the background from the modified ZYA1 method, and the background from the RPF method.  (See text for details.)  The data for all angles relative to the reaction plane have been scaled by 1/3 in the top panel in order to fit on the same scale.  The fit for the RPF method is to \Eref{Eqtn:JBBBCorrelationsRxnPlane} to order n = 3 from \dphirange\ {$\pi$/2} and has \chisq{151}{140}.  Bottom:  Ratios of the background from the RPF and ZYA1 methods to the true background.  }\label{Fig:PiecewiseFitBackgroundExtractionhhCentral}
\end{center}
\end{figure*}

\begin{figure*}
\begin{center}
\rotatebox{0}{\resizebox{17cm}{!}{
        \includegraphics{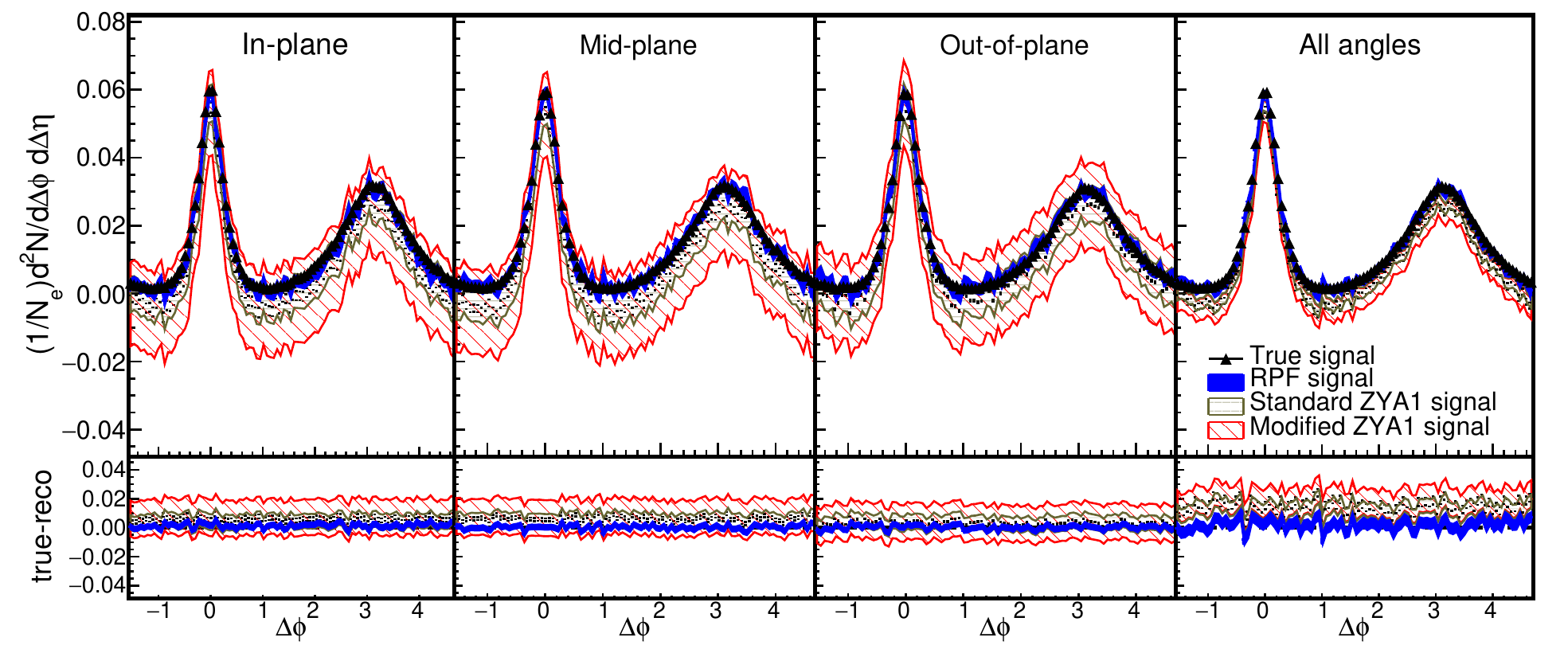}
}}
\caption{
Top:  The true signal for \dhcs in 0--10\% \Pb collisions at \sNN\ = 2.76 TeV for \inplane, \midplane, and \outplane triggers and for all triggers combined.  This is compared to the signal extracted using the background from the ZYA1 method, the modified ZYA1 method, and the background extracted from the RPF method for \dphirange\ {$\pi$/2} using the fit shown in \Fref{Fig:PiecewiseFitBackgroundExtractionhhCentral}.  (See text for details.)  The data for all angles relative to the reaction plane have been scaled by 1/3 in the top panel in order to fit on the same scale.  Bottom:  Differences between the true signal and the signal extracted using the background from the ZYA1 method, modified ZYA1 method, and the background from the RPF method.  }\label{Fig:PiecewiseFitTrueVsRecoSignalhhCentral}
\end{center}
\end{figure*}

The background determined using the ZYA1 method, modified ZYA1 method, and RPF method over the range \dphirange\ {$\pi$/2} are compared to the signal+background and the true background in our model for \dhcs in 0--10\% \Pb collisions at \sNN\ = 2.76 TeV for \inplane, \midplane, \outplane, and \allplane in \Fref{Fig:PiecewiseFitBackgroundExtractionhhCentral}.  The fit used in the RPF method is to order 
$n$ = 3 because the fit to order $n$ = 4 was not stable.  The reaction plane dependence of the correlations provide information on \vnumeff{2}, however, because \R{4} $\approx$ 0, there is no additional information on \vnumeff{4}.  \Fref{Fig:PiecewiseFitTrueVsRecoSignalhhCentral} compares the signal extracted with the ZYA1, modified ZYA1, and RPF methods and the yields are given in \Tref{Tab:Yields}.  This shows that the RPF method determines the signal with much higher accuracy and precision than the ZYA1 or modified ZYA1 methods.

\subsubsection{\Jhcs in 30-40\% central \Pb collisions with the RPF method over \dphirange\ {$\pi$/2}}

\begin{figure*}
\begin{center}
\rotatebox{0}{\resizebox{17cm}{!}{
        \includegraphics{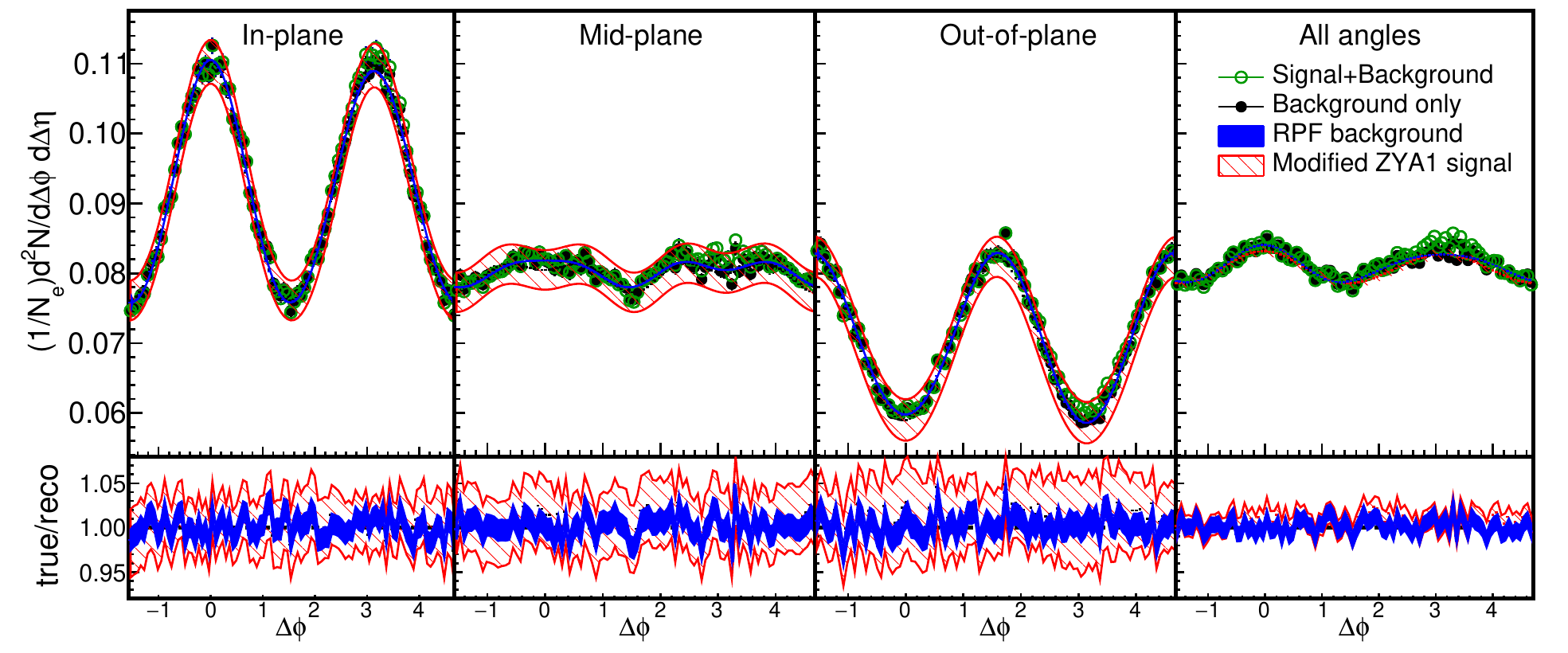}
}}\caption{
{
Top:  Signal+background for \jhcs in 30--40\% \Pb collisions at \sNN\ = 2.76 TeV in the region 1.0 $<|$\deta$|<$ 1.4 for \inplane, \midplane, and \outplane triggers and for all triggers combined.  This is compared to the true background, the background from the modified ZYA1 method, and the background from the RPF method.  (See text for details.)  The data for all angles relative to the reaction plane have been scaled by 1/3 in the tp panel in order to fit on the same scale.  The fit for the RPF method is to \Eref{Eqtn:JBBBCorrelationsRxnPlane} to order n = 4 from \dphirange\ {$\pi$/2} and has \chisq{193}{186}.  Bottom:  Ratios of the background from the RPF and ZYA1 methods to the true background.  }
}\label{Fig:PiecewiseFitBackgroundExtractionjethMidPeripheral}
\end{center}
\end{figure*}

\begin{figure*}
\begin{center}
\rotatebox{0}{\resizebox{17cm}{!}{
        \includegraphics{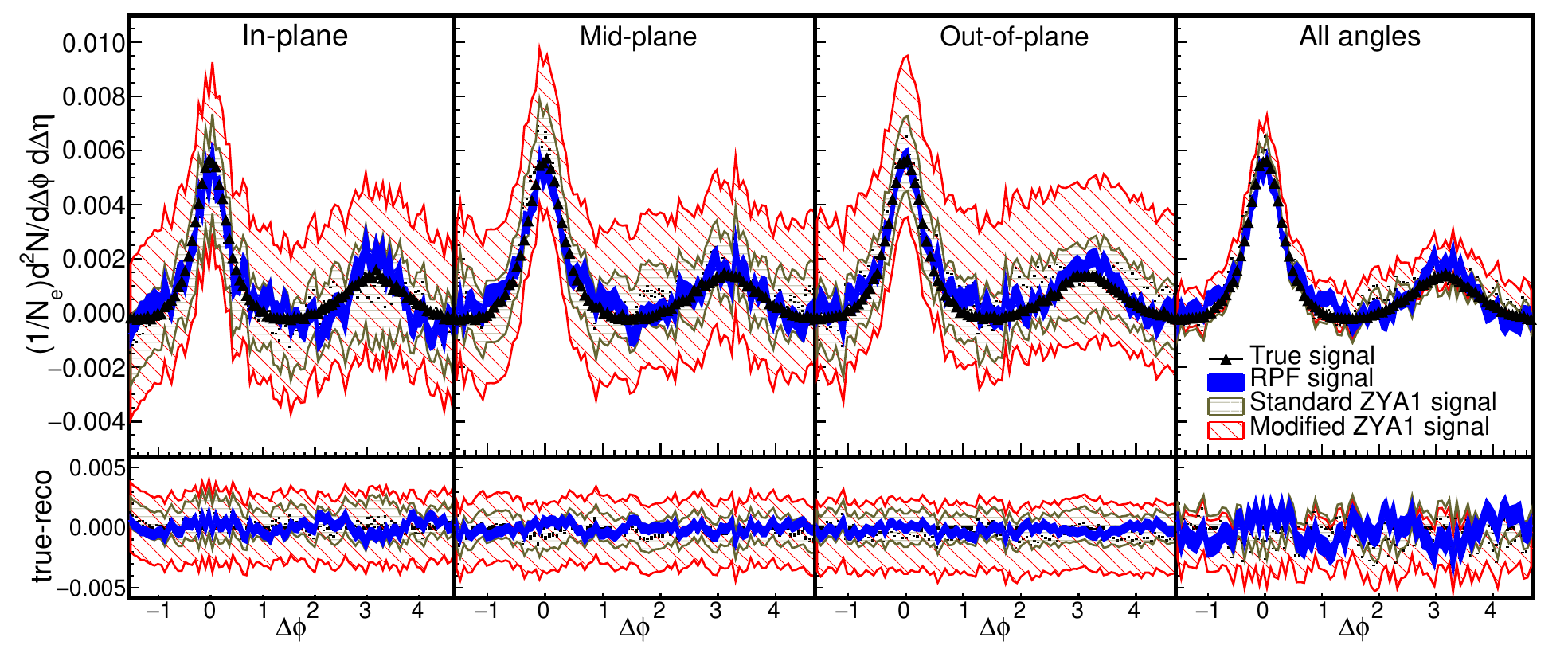}
}}
\caption{
Top:  The true signal for \jhcs in 30--40\% \Pb collisions at \sNN\ = 2.76 TeV for \inplane, \midplane, and \outplane triggers and for all triggers combined.  This is compared to the signal extracted using the background from the ZYA1 method, the modified ZYA1 method, and the background extracted from the RPF method for \dphirange\ {$\pi$/2} using the fit shown in \Fref{Fig:PiecewiseFitBackgroundExtractionjethMidPeripheral}.  (See text for details.)  The data for all angles relative to the reaction plane have been scaled by 1/3 in the top panel in order to fit on the same scale.  Bottom:  Differences between the true signal and the signal extracted using the background from the ZYA1 method, modified ZYA1 method, and the background from the RPF method.  }

\label{Fig:PiecewiseFitTrueVsRecoSignaljethMidPeripheral}
\end{center}
\end{figure*}

We also explore using the reaction plane dependence for determination of the background in \jhcs.  \Fref{Fig:PiecewiseFitBackgroundExtractionjethMidPeripheral} shows the signal+background in our model  in the region 1.0 $<|$\deta$|<$ 1.4 for \jhcs in 30--40\% \Pb collisions at \sNN\ = 2.76 TeV for \inplane, \midplane, \outplane, and \allplane and compares the background from the RPF method and the background from the modified ZYA1 method.  \Fref{Fig:PiecewiseFitTrueVsRecoSignaljethMidPeripheral} compares the true signal to the signal extracted from using the the RPF method, ZYA1, and modified ZYA1 methods to determine the background and the yields are given in \Tref{Tab:Yields}.  The RPF method works best.  This method would be particularly useful for \jhcs because only \vnum{2} has been measured for jets and therefore background subtraction requires either large estimates of the uncertainty due to the \vn~\cite{Adamczyk:2013jei} or limiting the analysis to the \ns where the background can be determines from the \deta dependence of the signal~\cite{CMS-PAS-HIN-14-016}.

\makeatletter{}\begin{table*}
\begin{center}
\caption{Yields $Y$ as defined in \Eref{Eq:Yields} scaled by 10$^{-3}$ from \Fref{Fig:PiecewiseFitTrueVsRecoSignalhhMidPeripheral}, \Fref{Fig:PiecewiseFitTrueVsRecoSignalhhMidPeripheralNarrowdPhi}, \Fref{Fig:PiecewiseFitTrueVsRecoSignalhhCentral}, and \Fref{Fig:PiecewiseFitTrueVsRecoSignaljethMidPeripheral}.  
For the true yield the statistical uncertainty is listed first followed by the uncertainty due to the subtraction of the background from the underlying event.  
For the ZYAM method the statistical uncertainty is listed first, followed by the uncertainty due to the background level and due to \vnumeffassoc{2} and \vnumefftrigger{2}.  
ZYA1 uncertainties are propagated assuming 100\% correlation between \vnumeffassoc{2} and \vnumefftrigger{2} and no correlation between the uncertainties on the level of the background and on the \vn.  Uncertainties on the ZYA1 method due to higher order \vn are not considered but are approximately 10\% of the uncertainties due to \vnumeffassoc{2} and \vnumefftrigger{2}.
}
\label{Tab:Yields}
\resizebox{\linewidth}{!}{ 
\begin{tabular}{c |c | c c c c | c c c c }
\hline
\hline
\multirow{2}{*}{ Sample }&  & \multicolumn{4}{c|}{\ns $Y\times 10^{-3}$} & \multicolumn{4}{c}{\as $Y\times 10^{-3}$} \\ 
 &  & \inplane & \midplane & \outplane & All & \inplane & \midplane & \outplane & All \\ [0.6ex]
\hline

& True  & 5.78 $\pm$ 0.03 $\pm$ 0.13 & 5.77 $\pm$ 0.03 $\pm$ 0.14 & 5.65 $\pm$ 0.03 $\pm$ 0.13 & 17.1 $\pm$ 0.1 $\pm$ 0.2 & 6.74 $\pm$ 0.03 $\pm$ 0.13 & 6.72 $\pm$ 0.03 $\pm$ 0.14 & 6.52 $\pm$ 0.03 $\pm$ 0.13 & 19.9 $\pm$ 0.1 $\pm$ 0.2\\
30--40\% & Mod. ZYA1  & 6.3 $\pm$ 5.9 $\pm$ 1.7 & 5.7 $\pm$ 6.0 $\pm$ 0.3 & 6.8 $\pm$ 6.1 $\pm$ 0.9 & 18.9 $\pm$ 4.2 $\pm$ 1.2 & 7.3 $\pm$ 5.9 $\pm$ 1.7 & 6.8 $\pm$ 6.0 $\pm$ 0.3 & 7.7 $\pm$ 6.1 $\pm$ 0.9 & 21.9 $\pm$ 4.2 $\pm$ 1.2\\
h-h & Std. ZYA1  & 4.5 $\pm$ 2.3 $\pm$ 1.7 & 5.5 $\pm$ 2.3 $\pm$ 0.3 & 5.6 $\pm$ 2.3 $\pm$ 0.9 & 15.7 $\pm$ 1.6 $\pm$ 1.2 & 5.5 $\pm$ 2.3 $\pm$ 1.7 & 6.5 $\pm$ 2.3 $\pm$ 0.3 & 6.5 $\pm$ 2.3 $\pm$ 0.9 & 18.7 $\pm$ 1.6 $\pm$ 1.2\\
& RPF (\dphirange{$\pi$/2})  & 5.5 $\pm$ 0.4 & 5.7 $\pm$ 0.3 & 5.9 $\pm$ 0.3 & 17.0 $\pm$ 0.7 & 6.6 $\pm$ 0.4 & 6.8 $\pm$ 0.3 & 6.8 $\pm$ 0.3 & 20.1 $\pm$ 0.7\\
& RPF (\dphirange{1})   & 5.7 $\pm$ 0.4 & 5.8 $\pm$ 0.4 & 5.9 $\pm$ 0.3 & 17.4 $\pm$ 0.7 & 6.8 $\pm$ 0.4 & 6.8 $\pm$ 0.4 & 6.8 $\pm$ 0.3 & 20.4 $\pm$ 0.7\\ [0.5ex]

\hline
& True  & 38.4 $\pm$ 0.1 $\pm$ 0.5 & 38.2 $\pm$ 0.1 $\pm$ 0.5 & 37.8 $\pm$ 0.1 $\pm$ 0.5 & 114.4 $\pm$ 0.2 $\pm$ 0.8 & 44.8 $\pm$ 0.1 $\pm$ 0.5 & 44.4 $\pm$ 0.1 $\pm$ 0.5 & 43.8 $\pm$ 0.1 $\pm$ 0.5 & 132.8 $\pm$ 0.2 $\pm$ 0.8\\
0--10\% & Mod. ZYA1  & 23 $\pm$ 26 $\pm$  3 & 23.2 $\pm$ 25.8 $\pm$ 0.3 & 29 $\pm$ 26 $\pm$  2 & 75.5 $\pm$ 18.3 $\pm$ 0.9 & 30 $\pm$ 26 $\pm$  3 & 29.9 $\pm$ 25.8 $\pm$ 0.3 & 36 $\pm$ 26 $\pm$  2 & 95.7 $\pm$ 18.3 $\pm$ 0.9\\
h-h & Std. ZYA1  & 29.2 $\pm$ 9.7 $\pm$ 2.8 & 27.6 $\pm$ 9.8 $\pm$ 0.3 & 29.8 $\pm$ 9.8 $\pm$ 2.3 & 86.7 $\pm$ 7.0 $\pm$ 0.9 & 35.4 $\pm$ 9.7 $\pm$ 2.8 & 34.4 $\pm$ 9.8 $\pm$ 0.3 & 37.1 $\pm$ 9.8 $\pm$ 2.3 & 106.9 $\pm$ 7.0 $\pm$ 0.9\\
& RPF (\dphirange{$\pi$/2})  & 35.5 $\pm$ 1.2 & 37.7 $\pm$ 0.9 & 36.0 $\pm$ 1.1 & 109.2 $\pm$ 2.5 & 41.7 $\pm$ 1.2 & 44.5 $\pm$ 0.9 & 43.3 $\pm$ 1.1 & 129.4 $\pm$ 2.6\\  [0.5ex]

 \hline
& True  & 4.41 $\pm$ 0.02 $\pm$ 0.10 & 4.43 $\pm$ 0.02 $\pm$ 0.10 & 4.43 $\pm$ 0.02 $\pm$ 0.10 & 13.19 $\pm$ 0.04 $\pm$ 0.17 & 1.67 $\pm$ 0.02 $\pm$ 0.10 & 1.67 $\pm$ 0.02 $\pm$ 0.10 & 1.70 $\pm$ 0.02 $\pm$ 0.10 & 4.96 $\pm$ 0.04 $\pm$ 0.17\\
30--40\% & Mod. ZYA1  & 4.7 $\pm$ 5.9 $\pm$ 1.7 & 5.2 $\pm$ 5.9 $\pm$ 0.3 & 6.1 $\pm$ 6.0 $\pm$ 0.9 & 16.2 $\pm$ 4.2 $\pm$ 1.2 & 2.0 $\pm$ 5.9 $\pm$ 1.7 & 2.5 $\pm$ 5.9 $\pm$ 0.3 & 3.4 $\pm$ 6.0 $\pm$ 0.9 & 8.2 $\pm$ 4.2 $\pm$ 1.2\\
jet-h & Std. ZYA1  & 3.6 $\pm$ 2.2 $\pm$ 1.7 & 4.8 $\pm$ 2.3 $\pm$ 0.3 & 4.6 $\pm$ 2.3 $\pm$ 0.9 & 13.2 $\pm$ 1.6 $\pm$ 1.2 & 0.9 $\pm$ 2.2 $\pm$ 1.7 & 2.2 $\pm$ 2.3 $\pm$ 0.3 & 1.9 $\pm$ 2.3 $\pm$ 0.9 & 5.2 $\pm$ 1.6 $\pm$ 1.2\\
& RPF (\dphirange{$\pi$/2})  & 4.3 $\pm$ 0.4 & 4.6 $\pm$ 0.3 & 4.8 $\pm$ 0.3 & 13.5 $\pm$ 0.6 & 1.6 $\pm$ 0.4 & 1.9 $\pm$ 0.3 & 2.1 $\pm$ 0.3 & 5.5 $\pm$ 0.6\\

\hline
\hline
\end{tabular}
}
\end{center}
\end{table*}

\section{Conclusions}\label{Sec:Conclusions}
 \makeatletter{}We have presented two methods for determining the combinatorial background in \dhcs and \jhcs by fitting the raw correlation at large \deta where the correlation is background-dominated.  We have demonstrated that these methods accurately and reliably subtract the background using a model where the background is entirely due to flow and the signal is generated using PYTHIA.  These methods produces better results than the ZYA1 method.  The RPF method is more accurate than ZYA1 even when the \dphi range of the fit is restricted and even in central collisions where the reaction plane resolution is poor.  These methods will be particularly useful for \jhcs since only \vnum{2} has been measured for jets, limiting the application of methods such as ZYAM, ZYA1, or the ABS method~\cite{Sickles:2009ka} which all require the \vn as input.  The reaction plane fit method makes the same assumptions about the shape of the background made in other methods, namely that it has the functional form given by \Eref{Eqtn:JBBBCorrelations}.  However, it does not make assumptions about the level of the background B or the \vneff.  The primary assumptions of this method are that the background has the functional form given by \Eref{Eqtn:JBBBCorrelations}, that the background's \deta dependence is negligible, and that the contribution of the signal to the correlation at large \deta and small \dphi is negligible.  This latter assumption is not valid at lower momenta and we were unable to extend the analysis to \ptassoc $<$ 1 \GeV because in this region the \ns peak is too broad even in PYTHIA and distorts the fit.  
We foresee future research using a two-dimensional fit to the near-side signal with a Gaussian in \dphi and \deta in order to extend the analysis to lower momenta. 
 \section{Acknowledgements}
 \makeatletter{}We are grateful to Rene Bellwied, Jana Biel\v{c}ikova, Marco van Leeuwen, and Soren Sorensen for useful comments on the manuscript and to Ken Read for helpful discussion on error analysis.  This work was supported in part by funding from the Division of Nuclear Physics of the U.S. Department of Energy under Grant No. DE-FG02-96ER40982. 

\makeatletter{}
    
\end{document}